\documentclass[journal]{IEEEtran}
\usepackage{amsmath,amssymb}
\usepackage{subfigure}
\usepackage{graphicx,graphics,color,psfrag}
\usepackage{cite,balance}
\usepackage{caption}
\allowdisplaybreaks
\usepackage{algorithm}
\usepackage{algorithmic}
\usepackage{accents}
\usepackage{amsthm}
\usepackage{bm}
\usepackage{url}
\usepackage[english]{babel}
\usepackage{multirow}
\usepackage{enumerate}
\usepackage{cases}
\usepackage{stfloats}
\usepackage{dsfont}
\usepackage{color,soul}
\usepackage{amsfonts}
\usepackage{cite,graphicx,amsmath,amssymb}
\usepackage{subfigure}
\usepackage{fancyhdr}
\usepackage{hhline}
\usepackage{graphicx,graphics}
\usepackage{array,color}
\usepackage{mathtools}
\usepackage{amsmath}

\newtheorem{corollary}{\emph{\underline{Corollary}}}

\newtheorem{proposition}{\emph{\underline{Proposition}}}

\newtheorem{example}{\bf Example}
\newtheorem{remark}{\bf \emph{\underline{Remark}}}

\def\({\left(}
\def\){\right)}

\def\b0{{\mathbf{0}}}

\newcommand{\nn}{\nonumber}

\begin{document}
\captionsetup[figure]{name={Fig.}}

\title{\huge 
Rotatable Antenna Enabled Multi-Cell Mixed Near-Field and Far-Field Communications} 
\author{Yunpu~Zhang,~\IEEEmembership{Graduate Student Member,~IEEE}, Changsheng~You,~\IEEEmembership{Member,~IEEE}, Ruichen~Zhang,  Beixiong~Zheng,~\IEEEmembership{Senior Member,~IEEE}, Hing Cheung So,~\IEEEmembership{Fellow,~IEEE}, \\Dusit Niyato,~\IEEEmembership{Fellow,~IEEE}, 
and Tony Q. S. Quek,~\IEEEmembership{Fellow,~IEEE}
\thanks{
\emph{(Corresponding author: Changsheng You.)}

Y. Zhang and H. C. So are with the Department of Electrical Engineering, City University of Hong Kong, Hong Kong (e-mail:
yunpu.zhang@my.cityu.edu.hk, hcso@ee.cityu.edu.hk). 

C. You is with the Department of Electronic and Electrical Engineering, Southern University of Science and Technology (SUSTech), Shenzhen
518055, China (e-mail: youcs@sustech.edu.cn).

R. Zhang and D. Niyato are with the College of Computing and
Data Science, Nanyang Technological University, Singapore 639798 (email: ruichen.zhang@ntu.edu.sg, dniyato@ntu.edu.sg). 

B. Zheng is with the School of Microelectronics,
South China University of Technology, Guangzhou 511442, China (e-mail: bxzheng@scut.edu.cn).

T. Q. S. Quek is with the Information System Technology
and Design Pillar, Singapore University of Technology and Design, Singapore
487372 (e-mail: tonyquek@sutd.edu.sg).}}
\maketitle
\begin{abstract}
    Prior studies on mixed near-field and far-field communications have focused exclusively on \emph{single-cell} scenarios, where both near-field and far-field users are served by the same base station (BS), leading to \emph{intra-cell} mixed-field interference. In this paper, we consider a more general and practical \emph{multi-cell mixed-field} scenario consisting of multiple cells, each serving multiple users, thus resulting in more complex \emph{inter-cell} mixed-field interference. To address this new challenge, we propose leveraging \emph{rotatable antenna} (RA) technology to enhance multi-cell mixed-field communication performance by exploiting the additional spatial degree-of-freedom (DoF) introduced by RA rotation to mitigate interference in an efficient way. Specifically, we study an RA-enabled multi-cell mixed-field communication system in which each BS is equipped with an RA array to serve its associated users. We formulate a network-wide sum-rate maximization problem that jointly optimizes the transmit beamforming and the rotation angles of the RA arrays, subject to per-BS power constraints and admissible array rotation limits. To gain useful insights into the role of RAs in multi-cell mixed-field communications, we first analyze a special case with a single user per cell. For this case, we obtain a closed-form expression for the \emph{rotation-aware inter-cell} mixed-field interference using the Fresnel integrals and analytically show that RA rotation can effectively mitigate such interference, thereby substantially improving system performance. For the general case with multiple users per cell, we develop an efficient \emph{double-layer} algorithm: the inner layer optimizes the transmit beamforming at each BS via semidefinite relaxation (SDR) and successive convex approximation (SCA); while the outer layer determines the rotation angles of the RA arrays using particle swarm optimization (PSO). Numerical results demonstrate that RA-enabled multi-cell systems achieve significant performance gains over conventional fixed-antenna systems, and the proposed joint design consistently outperforms various benchmark schemes.
    

\end{abstract}

\section{Introduction}
Extremely large-scale arrays (XL-arrays) are envisioned as a key enabler for future wireless systems, offering significant improvement in both
spectral efficiency and spatial resolution \cite{10496996,liu2023near,9903389,you2024next,10819473,11351318}. Notably, the large aperture of XL-arrays leads to a fundamental change in electromagnetic (EM) propagation
characteristics, causing a substantial transition from conventional far-field communications (planar wavefronts) to near-field communications characterized by spherical wavefronts. This thus leads to a more general and practical scenario: \emph{mixed
near-field and far-field communications}, where both far-field and near-field users coexist \cite{zhang2023mixed}. The distinct propagation characteristics of the near-field and far-field regimes give rise to a unique \emph{energy-spread} effect. Specifically, when a far-field beam is steered toward a specific angle, its energy may spread across
multiple directions in the near field. This effect may result in severe mixed-field inter-user interference in information transmission \cite{zhang2023mixed}, and mixed-field information leakage in secure communications \cite{liu2025physical}, both of which have been examined in single-cell scenarios. In this paper, we consider a more general scenario, namely, \emph{multi-cell mixed near-field and far-field communications}. In particular, we demonstrate that multi-cell mixed-field communications introduce even more complex \emph{intra-cell} near-field interference and \emph{inter-cell} mixed-field interference, which, however, can be effectively mitigated by leveraging rotatable antenna (RA) technology.

\subsection{Prior Works}
Existing research on mixed near-field and far-field communications has focused exclusively on single-cell scenarios/applications, where both near-field and far-field users are assumed to be located within the same cell \cite{11196008,wu2025scalable,zhang2023mixed,liu2025physical,zhang2023swipt}. The most unique characteristic of mixed-field communications is the so-called \emph{energy-spread} effect, which introduces both challenges and opportunities for system design.
In particular, \cite{zhang2023mixed} was the first to reveal the complicated interference characteristics in mixed-field communications arising from this effect, showing that a near-field user may experience strong interference from the beam directed toward a far-field user even when the spatial angle of the near-field user is in the vicinity of that of the far-field user. 
Furthermore, \cite{liu2025physical} investigated physical layer security in environments with near-field eavesdroppers and far-field legitimate users, and demonstrated that the information leakage from other legitimate users can be effectively exploited to weaken the eavesdropper’s ability to intercept the target user’s signal by leveraging the energy-spread effect.
On the other hand, the energy-spread effect can also be beneficial for system designs. 
For example, in \cite{zhang2023swipt}, the authors studied simultaneous wireless information and power transfer in mixed-field channels, where energy harvesting (EH) receivers are located in the near field and information receivers in the far field. It was revealed that the far-field beams can be harnessed to charge neighboring near-field EH receivers.

Movable antenna (MA) and fluid antenna have demonstrated superior capabilities in interference suppression \cite{Zhu_Corre,9650760}, but their complex positional parameters offer limited analytical insights. As a simplified implementation of the six-dimensional movable antenna (6DMA) \cite{11142311,shao6d,shao20246d,10883029}, RA technology has attracted growing interest due to its low implementation cost and reduced complexity \cite{zheng2025rotatableM,zheng2025rotatableJ,zheng2025rotatablePLS}. 
In particular, RAs can rotate the antenna radiation pattern in the angular domain to achieve flexible beam coverage through the adjustment of their rotation angles. Initial studies have showcased the great potential of RA-enabled communication systems \cite{zheng2025rotatableJ,zheng2025rotatablePLS,zhou2025rotatable,rotationISAC}. Specifically, in \cite{zheng2025rotatableJ}, the authors established the system model and channel characterization for RA-enabled systems and theoretically demonstrated the performance gains brought by RA. In \cite{zheng2025rotatablePLS}, the authors designed an RA-enabled PLS system with a single legitimate user and multiple eavesdroppers. It was shown that the RA array can significantly enhance the array gain toward the legitimate user while simultaneously suppressing signal leakage to the eavesdroppers. Furthermore, \cite{zhou2025rotatable} integrated the RA array into an integrated sensing and communications (ISAC) framework, analytically demonstrating improvements in both sensing and communication aspects. In \cite{rotationISAC}, the authors employed RAs to enable cooperative ISAC among multiple base stations (BSs), where each BS is equipped with an RA array to optimize the sensing beampattern matching while maintaining the communication performance.

\subsection{Motivations and Contributions}
The effectiveness of RAs in mitigating both near-field and mixed-field interference in mixed-field communications has been theoretically revealed in our previous work \cite{zhangRAmixed}, which, however, focused solely on the single-cell case where both near-field and far-field users are served by the same BS. In contrast, under the XL-array setting, \emph{multi-cell mixed-field communications} naturally arise for two key reasons. First, with XL-arrays, users in a given cell are more likely to be located in the near-field region of their serving BS. For instance, an XL-array BS with an aperture of $1$ m operating at $28$ GHz has a Rayleigh distance of approximately $187$ m. This implies that, in a typical cellular network with a cell radius of $200$ m, most users within the cell will reside in the near-field region of their serving BS \cite{liu2025physical}. Second, these users may also lie in the far-field region of BSs in neighboring cells.

For conventional single-cell mixed-field systems, the system needs to deal with both \emph{intra-cell} near-field interference and \emph{intra-cell} mixed-field interference, both of which arise from the simultaneous transmissions of multiple users located in the same cell and served by the same BS. In contrast, multi-cell mixed-field communications are subject to \emph{intra-cell} near-field interference as well as \emph{inter-cell} mixed-field interference, where the latter is caused by transmissions from users or BSs in neighboring cells. The coexistence of these interference sources results in more intricate interference patterns and, if not properly mitigated, can lead to significant performance degradation. To the best of our knowledge, such interference characteristics in multi-cell mixed-field communications have not yet been well studied in the literature, thus underscoring the need for dedicated transmission designs to achieve their effective mitigation and multi-cell communication performance enhancement. Moreover, the role of RAs in multi-cell mixed-field communications has not been theoretically analyzed, and their potential to tackle the corresponding interference challenges inherent in such systems remains largely unexplored.
 

\begin{figure}[t]
	\centering
\includegraphics[width=0.45\textwidth]{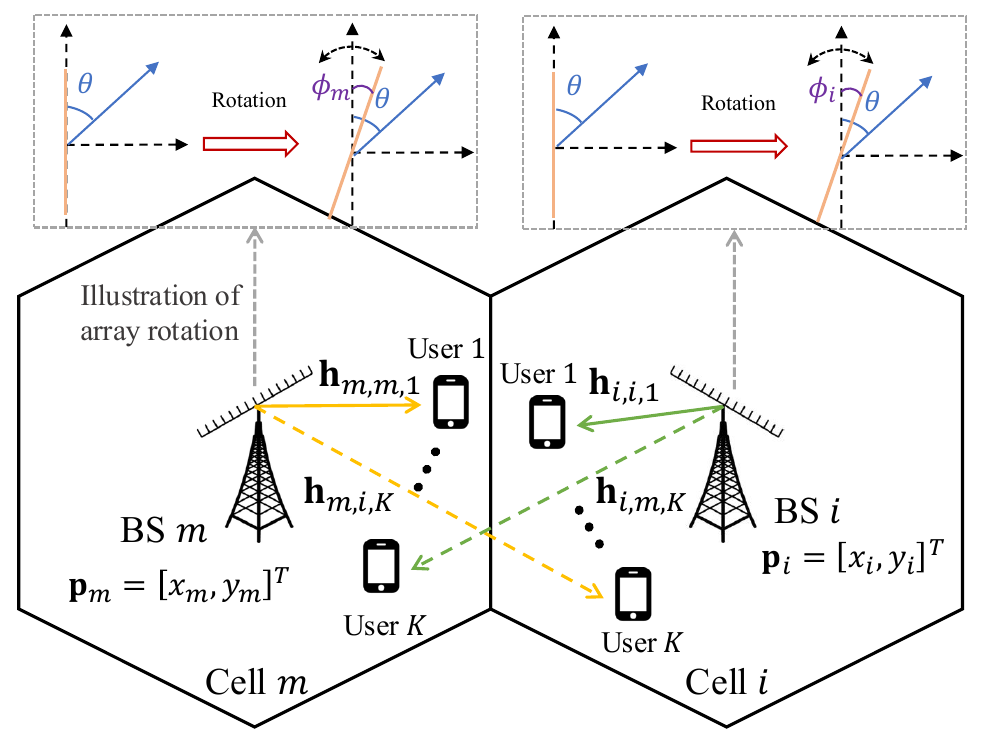}
	\caption{{Illustration of RA-enabled multi-cell mixed-field communication
system.}} \label{Fig:SM}
\end{figure}

Motivated by the above, we consider in this paper a new and practical RA-enabled \emph{multi-cell mixed near-field and far-field} communication system, aiming to exploit the performance gains provided by RA rotation. To this end, we propose equipping the BS in each cell with an RA array, as illustrated in Fig. \ref{Fig:SM}, and focus on joint optimization of the transmit beamforming and the RA-array rotation angle across cells to maximize the achievable sum-rate of all users in the multi-cell system. 
Our main contributions are summarized as follows.
\begin{itemize}
    \item First, this work represents the \emph{first} attempt to study the RA-enabled transceiver design in a new multi-cell mixed near-field and far-field communication system.
    Specifically, we propose equipping each BS with an RA array to enhance the performance of multi-cell mixed-field communications. An optimization problem is then formulated to maximize the sum-rate of all users across multiple cells by jointly optimizing the transmit beamforming and the RA-array rotation angle at each BS, subject to the per-cell power budgets and the admissible rotation constraints.
    \item Second, to provide fundamental insights into the role of RAs in multi-cell mixed-field communication systems, we analyze a special case where each cell serves a single user. In particular, we obtain a closed-form expression for the \emph{rotation-aware} \emph{inter-cell} mixed-field interference using the Fresnel integrals, and identify the key factors that determine such interference. Our results highlight a sharp contrast with conventional single-cell scenarios regarding these factors. We further analytically demonstrate that array rotation effectively suppresses the \emph{inter-cell} mixed-field interference, thereby significantly enhancing multi-cell mixed-field communication performance.
    \item Third, for the general case where each cell serves multiple users, the resulting problem is a challenging non-convex optimization problem with highly coupled variables. To address this challenge,
    we develop an efficient \emph{double-layer} algorithm to obtain a high-quality solution. Our method consists of two layers: the inner layer optimizes the transmit beamforming via the semidefinite relaxation (SDR) and successive convex approximation (SCA), while the outer layer determines the RA-array rotation angles across cells using the particle swarm optimization (PSO). 
    \item   Finally, numerical results are provided to illustrate the substantial performance gains achieved by incorporating RAs into multi-cell mixed-field systems and confirm the effectiveness of the proposed joint optimization approach. In particular, our findings include: 1) RA-enabled systems achieve a much higher performance gain than fixed-antenna schemes in multi-cell mixed-field scenarios; and 2) the proposed joint design consistently outperforms various benchmarks that optimize either component independently.
\end{itemize}
\subsection{Organization and Notations}
\subsubsection{Organization} The remainder of this paper is structured as follows. Section \ref{Sec:SM} introduces the system model and problem formulation for the multi-cell mixed-field communication scenario. Section \ref{Sec:analytical} analyzes a special case where each cell serves a single user, providing key insights into the impact of RAs on mitigating the inter-cell interference. Section \ref{Sec:Proposed} addresses the general case by proposing a double-layer algorithm to obtain a high-quality solution. Section \ref{Sec:NR} presents numerical results to evaluate the performance of the developed joint design and demonstrates the superiority of RA-enabled multi-cell systems over conventional fixed-antenna counterparts. Finally, Section \ref{Sec:con} concludes the paper.

 \subsubsection{Notations}
 Bold lowercase letters denote vectors, bold uppercase letters denote matrices, and calligraphic uppercase letters denote sets.  The operators $(\cdot)^T$ and $(\cdot)^H$ represent the transpose and Hermitian transpose, respectively.
A complex Gaussian random variable with mean $\mu$ and variance $\sigma^2$ is expressed as $\mathcal{CN}(\mu,\sigma^2)$.
The notation $|\cdot|$ indicates absolute value for scalars and cardinality for sets. For a vector $\mathbf{x}$, $[\mathbf{x}]_i$ denotes its $i$-th element, while for a matrix $\mathbf{X}$, $\mathrm{Tr}(\mathbf{X})$ and $\mathrm{Rank}(\mathbf{X})$ denote its trace and rank, respectively, and $\mathbf{X} \succeq 0$ denotes that $\mathbf{X}$ is positive semidefinite. The Frobenius norm is denoted by $\|\cdot\|_F$, and big-O notation $\mathcal{O}(\cdot)$ is used for computational complexity.

\section{System Model and Problem Formulation}\label{Sec:SM}
We consider an RA-enabled mixed near-field and far-field multi-cell downlink communication system, as depicted in Fig. \ref{Fig:SM}. The system consists of $M$ macro cells, indexed by $\mathcal{M}=\{1,\ldots,M\}$, where we assume each cell contains a BS (e.g., the BS in cell $m$) serving $K$ single-antenna users indexed by $\mathcal{K}_{m}=\{1,\ldots,K\}$. In particular, each BS is equipped with a uniform linear RA array\footnote{The analytical results and the proposed design can be extended to a uniform planar array (UPA) by constructing the near-field and far-field steering vectors via Kronecker products of the corresponding one-dimensional steering vectors in \eqref{Eq:near-field} and \eqref{Eq:far-field}. In this case, the steering vectors depend on both azimuth and elevation angles, and the effects of array rotation as well as the associated design can be generalized by considering three-dimensional rotations in the azimuth and elevation domains.} consisting of $N=2\tilde{N}+1$ antennas with half-wavelength spacing. The reference coordinate of BS $m$ is denoted by $\mathbf{p}_{m}=[x_m,y_m]^T$. In addition to the transmit beamforming design at each BS, the one-dimensional rotation angles of the RA arrays, represented by $\boldsymbol{\phi}=[\phi_1,\ldots,\phi_M]^T$, can be adjusted to mitigate complex intra-cell near-field interference and inter-cell mixed-field interference, thereby enhancing overall system performance. 

\subsection{Mixed Near-Field and Far-Field Channel Models}
To characterize the boundary between the near-field and far-field regions, we adopt the \emph{effective Rayleigh distance}, given by $R_{\rm Ray}(\theta)=\upsilon\sin^2{\theta}\frac{2D^2}{\lambda}$, where $\upsilon=0.367$ \cite{cuiwideband}, $\theta$ is the user angle, $D$ is the XL-array aperture, and $\lambda$ denotes the carrier wavelength.
In multi-cell communication scenarios with XL-arrays, near-field and far-field transmissions inherently coexist. To capture this effect, we assume that users within a given cell lie in the near-field region of their serving BS while falling into the far-field region of neighboring BSs.\footnote{When the cell radius is less than or equal to the effective Rayleigh distance, the analysis remains valid. If the radius exceeds this distance, some edge users may experience inter-cell near-field interference, as discussed in \cite{zhangRAmixed}. Nevertheless, the majority of users still conform to the inter-cell mixed-field interference model analyzed in this paper.} Based on this setup, we present the near-field and far-field channel models in the following.\footnote{In this work, mixed-field propagation is characterized through the joint consideration of near-field and far-field channel models. It is worth noting that the distinctive interference characteristics commonly associated with mixed-field communications are not an artifact of heterogeneous near-field and far-field channel modeling. Even when all users are modeled using near-field channels, similar mixed-field interference behavior naturally arises in practical multi-cell deployments, since users are typically located in the near field of their serving BS while being effectively in the far field of neighboring BSs due to much larger inter-site distances. The mixed-field formulation enables clearer closed-form interference expressions in \eqref{Eq:rho_unified_approx}, thereby facilitating analytical insights into the impact of array rotation on interference mitigation. }

\subsubsection{Intra-cell near-field channel} 
For a typical near-field user within each cell, the channel from its serving BS to this user is modeled using spherical-wavefront propagation. Specifically, consider user $k$ in cell $m$, which is located at range $r_{m,k}$ and intra-cell angle $\theta_{m,k}$ from the center of the $m$-th RA array. The distance between user $k$ and the $n$-th antenna is then expressed as
\begin{align}
   &r^{(n)}_{m,k} =  \sqrt{r^2_{m,k}+n^2d^2-2r_{m,k}\cos{(\phi_m-\theta_{m,k})}},\nn\\
    & \overset{(a)}{\approx }\!r_{m,k}\!-\!nd\cos{(\phi_m-\theta_{m,k})}\!+\!\frac{n^2d^2\sin^2{(\phi_m\!-\!\theta_{m,k})}}{2r_{m,k}},
\end{align}
where $(a)$ is obtained using the second-order Taylor expansion, i.e., $\sqrt{1+x}\approx 1+\frac{1}{2}x-\frac{1}{8}x^2$ \cite{zhang2023mixed}, and $\phi_m$ denotes the rotation angle of the RA array at the $m$-th BS. In this paper, we adopt the general multipath channel model for all communication users, which consists of a line-of-sight (LoS) path and multiple non-LoS (NLoS) paths caused by environmental scatterers.\footnote{We assume that each BS has the perfect channel state information (CSI) for all near-field and far-field channels in the multi-cell mixed-field system. This assumption facilitates the subsequent analytical insights into the impact of array rotation on mixed-field interference, as well as the characterization of the performance upper bound of multi-cell mixed-field communication systems. }
Accordingly, the near-field channel from the $m$-BS to the $k$-th user in the $m$-th cell is given by
\begin{align}
    \mathbf{h}^{H}_{m,m,k} &= \sqrt{N}\beta_{m,k}\mathbf{b}^H(\theta_{m,k},r_{m,k},\phi_m)\nn\\
    &+\sqrt{\frac{N}{L_{m,k}}}\sum_{\ell=1}^{L_{m,k}}\beta_{m,k,\ell}\mathbf{b}^H(\theta_{m,k,\ell},r_{m,k,\ell},\phi_m),
\end{align}
where $\beta_{m,k}$ and $\beta_{m,k,\ell}$ denote the complex gains of the LoS path and the $\ell$-th NLoS path associated with user $k$ in cell $m$, respectively.\footnote{For analytical simplicity, we consider omnidirectional antenna gain model in this paper, whereas the proposed scheme can be extended to the general directional antenna gain model \cite{zheng2025rotatableJ}. } The parameters $\theta_{m,k,\ell}$ and $r_{m,k,\ell}$ denote the intra-cell angle and range of the $\ell$-th scatterer associated with $k$-th user in cell $m$ with respect to the center of the RA array of the $m$-th BS, respectively.
The near-field steering vector $\mathbf{b}(\theta_{m,k},r_{m,k},\phi_m) \in \mathbb{C}^{N\times1}$, which  incorporates the effect of RA-array rotation at the $m$-th BS, is given by
\begin{align}\label{Eq:near-field}
    \mathbf{b}^H(\theta_{m,k},r_{m,k},\phi_m) = \frac{1}{\sqrt{N}}\Big[&e^{-j\frac{2\pi}{\lambda}(r^{(-\Tilde{N})}_{m,k}-r_{m,k})},\ldots,\nn\\
    &e^{-j\frac{2\pi}{\lambda}(r^{(\Tilde{N})}_{m,k}-r_{m,k})}\Big].
\end{align}

 \begin{figure}[t]
	\centering
\includegraphics[width=0.46\textwidth]{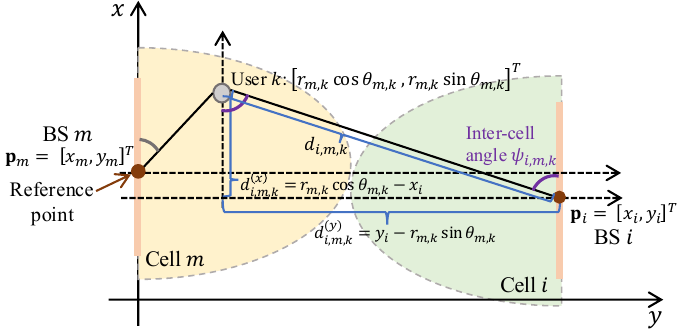}
	\caption{{Illustration of inter-cell angle determination.}} \label{Fig:equivalent}
\end{figure}

\subsubsection{Inter-cell far-field channel} In contrast, the channel between the $k$-th user in cell $m$ and a neighboring BS $i \in \mathcal{M}\setminus \{m\}$ is modeled using the far-field planar-wavefront propagation and is thus given by
\begin{align}
    \mathbf{h}^{H}_{i,m,k} = &\sqrt{N}\tilde{\beta}_{i,k}\mathbf{a}^H(\psi_{i,m,k},\phi_i)\nn\\
    &+\sqrt{\frac{N}{L_{i,m,k}}}\sum^{L_{i,m,k}}_{\ell=1}\tilde{\beta}_{i,k,\ell}\mathbf{a}^H(\psi_{i,m,k,\ell},\phi_i),
\end{align}
where $\tilde{\beta}_{i,k}$ and $\tilde{\beta}_{i,k,\ell}$ denote the complex channel gains of the LoS path and the $\ell$-th NLoS path, respectively, between BS $i$ and user $k$ in cell $m$.
As illustrated in Fig. \ref{Fig:equivalent}, let $\psi_{i,m,k}$ denote the \emph{inter-cell angle} of user $k$ in cell $m$ relative to the BS in cell $i$, i.e., BS $i$. The
far-field channel steering vector $\mathbf{a}(\psi_{i,m,k},\phi_i) \in \mathbb{C}^{N\times1}$, accounting for the RA-array rotation of BS $i$, is given by
\begin{align}\label{Eq:far-field}
    \mathbf{a}^H(\psi_{i,m,k},\phi_i) = \frac{1}{\sqrt{N}}\Big[&e^{j\frac{2\pi}{\lambda}\Tilde{N}d\cos{(\psi_{i,m,k}-\phi_i)}},\ldots,\nn\\
    &e^{j\frac{2\pi}{\lambda}-\Tilde{N}d\cos{(\psi_{i,m,k}-\phi_i)}}\Big].
\end{align}
 The inter-cell angle $\psi_{i,m,k}$ is determined via the triangle inequality, computed as
\begin{equation}
    \psi_{i,m,k}= 
    \begin{cases}
      \arctan \left(\frac{d^{(y)}_{i,m,k}}{d^{(x)}_{i,m,k}}\right),&\text{if}~ d^{(x)}_{i,m,k}\ge 0, \\
      \arctan \left(\frac{d^{(y)}_{i,m,k}}{d^{(x)}_{i,m,k}}\right)+\pi, &\text{otherwise},
    \end{cases}
\end{equation}
where $d^{(x)}_{i,m,k}=r_{m,k}\cos\theta_{m,k}-x_i$ and $d^{(y)}_{i,m,k}=y_i-r_{m,k}\sin\theta_{m,k}$ denote the $x$- and $y$-axis projections of user $k$ in cell $m$ with respect to BS $i$. Accordingly,
the inter-cell distance between user $k$ in cell $m$ and BS $i$ is calculated as 
\begin{equation}
   \!\! d_{i,m,k} \!= \!\sqrt{(r_{m,k}\cos\theta_{m,k}-x_i)^2\!+\!(y_i-r_{m,k}\sin\theta_{m,k})^2}.
\end{equation}

\subsection{Signal Model}
To reduce the high hardware cost associated with XL-arrays, we adopt a cost-effective hybrid beamforming architecture \cite{liu2025physical}, for which each BS is equipped with $K$ $(K\ll N)$ radio frequency (RF) chains to simultaneously serve $K$ users inside its serving cell. Let $\mathbf{F}_{{\rm A},m} \in \mathbb{C}^{N \times K}$ and $\mathbf{F}_{{\rm D},m} \in \mathbb{C}^{K \times K}$ denote the analog and digital beamformers of the $m$-th BS, respectively. Accordingly, the received signal at the $k$-th user in cell $m$ can then be expressed as
\begin{align}
    y_{m,k} = &\mathbf{h}^{H}_{m,m,k}\mathbf{F}_{{\rm A},m}\mathbf{f}_{{\rm D},m,k}s_{m,k}\nn\\
    &
    +\underbrace{\sum^{K}_{\ell=1,\ell\neq k}\mathbf{h}^{H}_{m,m,k}\mathbf{F}_{{\rm A},m}\mathbf{f}_{{\rm D},m,\ell}s_{m,\ell} }_{\text{intra-cell near-field interference} }\nn\\
    &+\underbrace{\sum^{M}_{i=1,i\neq m}\sum^{K}_{j=1}\mathbf{h}^{H}_{i,m,k}\mathbf{F}_{{\rm A},i}\mathbf{f}_{{\rm D},i,j}s_{i,j}}_{\text{inter-cell mixed-field interference} } + n_{m,k},
\end{align}
where $\mathbf{f}_{{\rm D},m,k}$ is the $k$-th column of $\mathbf{F}_{{\rm D},m}$. The symbol $s_{m,k}$ represents the signal transmitted by  BS $m$ to user $k$ in its cell, and $s_{m,\ell}$ denotes the signal sent by the same BS to another user $\ell\in\mathcal{K}_m,\ell\neq k$. Additionally, $s_{i,j}$ represents the signal transmitted by BS $i$ to user $j$ in cell $i$. The term $n_{m,k} \sim \mathcal{CN}(0,\sigma^2_{m,k})$ denotes the additive while Gaussian noise (AWGN) at user $k$ in cell $m$. As such, the achievable rate in bits per second per Hertz (bps/Hz) at user $k$ in cell $m$ is given by \eqref{Eq:userrate}, as shown at the top of the next page.
\begin{figure*}[t]
    \begin{align}\label{Eq:userrate}
    \!\!\!R_{m,k}\left(\left\{\mathbf{F}_{{\rm A},m},\mathbf{F}_{{\rm D},m}\right\},\boldsymbol{\phi}\right) \!= \!\log_2\left(1+\frac{|\mathbf{h}^{H}_{m,m,k}\mathbf{F}_{{\rm A},m}\mathbf{f}_{{\rm D},m,k}|^2}{\sum^{K}_{\ell=1,\ell\neq k}|\mathbf{h}^{H}_{m,m,k}\mathbf{F}_{{\rm A},m}\mathbf{f}_{{\rm D},m,\ell}|^2+\sum^{M}_{i=1,i\neq m}\sum^{K}_{j=1}|\mathbf{h}^{H}_{i,m,k}\mathbf{F}_{{\rm A},i}\mathbf{f}_{{\rm D},i,j}|^2+\sigma^2_{m,k}}\right)
\end{align}
\hrulefill
\end{figure*}
 \begin{remark}(Key difference between multi-cell mixed-field communications
and their near-field and far-field counterparts)
     \emph{It is important to emphasize that the fundamental difference between the multi-cell mixed-field communications and conventional near-field or far-field systems lies in the complexity of interference.
     In near-field (or far-field) multi-cell systems, both intra-cell and inter-cell interference occur predominantly within the same domain, i.e., near-field \cite{10892231} (or far-field \cite{9090356}). By contrast, multi-cell mixed-field communications exhibit distinct interference patterns. Specifically, the intra-cell interference arises primarily in the near-field, while the inter-cell interference corresponds to the \emph{mixed-field interference}. This additional complexity in interference patterns significantly degrades the performance of multi-cell systems. Consequently, effective interference mitigation is crucial to enhancing system performance, and this constitutes the main focus of this work, wherein we exploit the additional DoF offered by array rotation to suppress such interference.}
 \end{remark}

\subsection{Problem Formulation}
We aim to maximize the achievable sum-rate of all users across the multi-cell system by jointly optimizing the hybrid beamforming matrices at each BS, $\{\mathbf{F}_{{\rm A},m}, \mathbf{F}_{{\rm D},m}\}$, and the rotation angles of all RA arrays, $\boldsymbol{\phi}$. The resultant problem is formulated as:
\begin{subequations}
	\begin{align}
		({\bf P1}):~\max_{\substack{\{\mathbf{F}_{{\rm A},m}, \mathbf{F}_{{\rm D},m}\},\boldsymbol{\phi} }}  ~&\sum^{M}_{m=1}\sum^{K}_{k=1} R_{m,k}\left(\left\{\mathbf{F}_{{\rm A},m},\mathbf{F}_{{\rm D},m}\right\},\boldsymbol{\phi}\right)
		\\
		\text{s.t.}~~
		&	\|\mathbf{F}_{{\rm A},m}\mathbf{F}_{{\rm D},m}\|^2_F\le P,\forall{m},\label{P1:pow_cons}\\
        & |\mathbf{F}_{{\rm A},m}(i,j)|=1,\forall{(i,j)}\in\mathcal{F},\label{P1:ana_norm}\\
        & [\boldsymbol{\phi}]_m \in \mathcal{C}_{\phi_m}, \forall{m},\label{P1:rot_angle}
	\end{align}
 \end{subequations}
where $P$ denotes the maximum transmit power per BS, \eqref{P1:ana_norm} imposes the unit-modulus constraint on the analog beamformer of BS $m$, with $\mathcal{F}$ representing the index set of its non-zero entries, and \eqref{P1:rot_angle} confines the RA rotation angle of each BS to its allowable rotation region $\mathcal{C}_{\phi_{m}}=\left[\phi_{m,{\rm min}},\phi_{m,{\rm max}}\right]$. 

\begin{figure}[t]
	\centering
\includegraphics[width=0.46\textwidth]{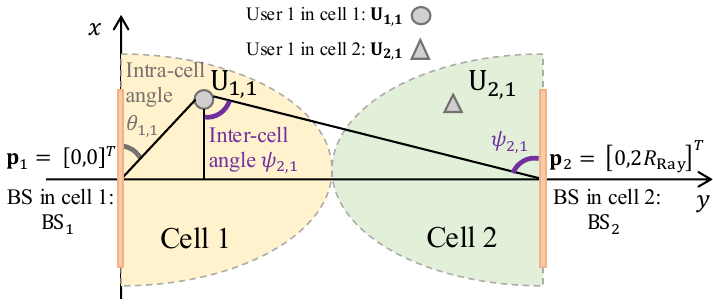}
	\caption{{Illustration of considered two-cell scenario.}} \label{Fig:two-cell-special}
\end{figure}
\section{Effect of Array Rotation on Inter-Cell Mixed-Field Interference}\label{Sec:analytical}
In this section, we first analytically characterize the effect of array rotation on the unique inter-cell mixed-field interference, and then reveal a fundamental trade-off in employing RA array rotation to mitigate intra-cell near-field interference and inter-cell mixed-field interference.
\subsection{Effect of Array Rotation on Mixed-Field Interference}\label{Sec:ana}
To provide key insights into the effect of array rotation on inter-cell mixed-field interference, we consider a special case depicted in Fig. \ref{Fig:two-cell-special}, where there are two cells ($M=2$) with two BSs located at $\mathbf{p}_1 = [0,0]^T$ and $\mathbf{p}_2 = [0,2R_{\rm Ray}]^T$, each serving a single user (i.e., $K=1$). For brevity, the users in the first and the second cells are denoted by ${\rm U}_{1,1}$ and ${\rm U}_{2,1}$, respectively, while the BSs are referred to as ${\rm BS}_1$ and ${\rm BS}_2$, respectively. The inter-cell angles of ${\rm U}_{1,1}$ and ${\rm U}_{2,1}$  with respect to ${\rm BS}_1$ and ${\rm BS}_2$ are denoted by $\psi_{2,1}$ and $\psi_{1,2}$, respectively.
The achievable sum-rate of ${\rm U}_{1,1}$ and ${\rm U}_{2,1}$ can be rewritten as
\begin{align}\label{sc:sumR1}
    R = &\log_2\left(1+\frac{|\mathbf{h}^H_{1,1,1}\mathbf{F}_{{\rm A},1}\mathbf{f}_{{\rm D},1,1}|^2}{|\mathbf{h}^H_{2,1,1}\mathbf{F}_{{\rm A},2}\mathbf{f}_{{\rm D},2,1}|^2+\sigma^2_{1,1}}\right)
    \nn\\
    &+\log_2\left(1+\frac{|\mathbf{h}^H_{2,2,1}\mathbf{F}_{{\rm A},2}\mathbf{f}_{{\rm D},2,1}|^2}{|\mathbf{h}^H_{1,2,1}\mathbf{F}_{{\rm A},1}\mathbf{f}_{{\rm D},1,1}|^2+\sigma^2_{2,1}}\right).
\end{align}

Since the considered scenario involves only one user per cell, the intra-cell near-field inter-user interference is absent. Nevertheless, it is worth noting that, in a multi-cell mixed-field system, the characteristics of this interference are consistent with those observed in conventional single-cell near-field inter-user interference. The effect of RA-array rotation on such interference has been theoretically studied in our previous work \cite{zhangRAmixed}. Moreover, the trade-off of array rotation in mitigating intra-cell near-field interference and inter-cell mixed-field interference will be evaluated in Section~\ref{Sec:tradoff}. Therefore, in this section, we focus on analytically characterizing the effect of RA-array rotation on the unique inter-cell mixed-field interference.
To this end, we investigate the LoS-dominant scenario for each user in high-frequency bands \cite{zhang2023swipt,liu2025physical} and adopt a low-complexity hybrid beamforming design.\footnote{The LoS-dominant assumption facilitates analytical insights, and the resulting conclusions remain effective in general multipath scenarios with dominant geometric characteristics, as the LoS-based analysis captures the underlying geometric structure governing the impact of array rotation.  Investigating scenarios with severely blocked LoS or rich scattering is left for future work.} In particular, the analog beamformer is implemented using maximal ratio transmission (MRT), whereby each BS steers its beam towards the served near-field user, while the power allocation is optimized in the digital beamforming stage.\footnote{Note that more sophisticated digital beamforming schemes, such as zero-forcing (ZF) or minimum mean-squared error (MMSE), typically complicate the theoretical analysis and may even render it intractable. Therefore, in our performance analysis, we focus on power allocation design, with the optimization of the digital beamforming matrix addressed separately in Section \ref{Sec:Proposed}.} Mathematically, the corresponding hybrid beamforming vectors are expressed as
\begin{align}
    \mathbf{F}_{{\rm A},1}\mathbf{f}_{{\rm D},1,1}&= \sqrt{P_{1,1}}\mathbf{b}(\theta_{1,1},r_{1,1},\phi_{1}),\label{eq:bf1}\\
     \mathbf{F}_{{\rm A},2}\mathbf{f}_{{\rm D},2,1}&=\sqrt{P_{2,1}}\mathbf{b}(\theta_{2,1},r_{2,1},\phi_{2}),\label{eq:bf2}
\end{align}
where $P_{1,1},P_{2,1}\in [0,P]$ denote the transmit power allocated to ${\rm U}_{1,1}$ and ${\rm U}_{2,1}$, respectively. After applying the composite beamformers in \eqref{eq:bf1} and \eqref{eq:bf2},
the achievable sum-rate of ${\rm U}_{1,1}$ and ${\rm U}_{2,1}$ in \eqref{sc:sumR1} can then be rewritten as
\begin{align}\label{Eq:updateR}
    R &=  \log_2\left(1+\frac{P_{1,1}N|\beta_{1,1}|^2}{P_{2,1}N|\tilde{\beta}_{2,1}|^2\rho^2(\psi_{2,1},\theta_{2,1},r_{2,1},\phi_2)+\sigma^2_{1,1}}\right)
    \nn\\
    &+\log_2\left(1+\frac{P_{2,1}N|\beta_{2,1}|^2}{P_{1,1}N|\tilde{\beta}_{1,2}|^2\rho^2(\psi_{1,2},\theta_{1,1},r_{1,1},\phi_1)+\sigma^2_{2,1}}\right).
\end{align}
\noindent Here, the two cross-correlation terms,  $\rho(\psi_{2,1},\theta_{2,1},r_{2,1},\phi_2)$ and $\rho(\psi_{1,2},\theta_{1,1},r_{1,1},\phi_1)$ that quantify the intensity of inter-cell mixed-field interference can be expressed in a unified form as
\begin{align}\label{Eq:rho_unified} &\rho(\psi_{i,k},\theta_{i,m},r_{i,m},\phi_i) = |\mathbf{a}^H(\psi_{i,k},\phi_i) \mathbf{b}(\theta_{i,m},r_{i,m},\phi_i)| \nonumber \\ &=\frac{1}{N}\bigg|\sum_{n=-\tilde{N}}^{\tilde{N}}\exp\bigg(j\frac{2\pi}{\lambda}\bigg[ n^2\frac{d^2\sin^2{(\phi_i-\theta_{i,m})}}{2r_{i,m}} \nonumber \\ &\quad\quad\quad+n(d\cos{(\psi_{i,k}-\phi_i)}- d\cos{(\phi_i-\theta_{i,m}))} \bigg]\bigg)\bigg|, \end{align}
where $(i,k,m)\in\{(2,1,1),(1,2,1)\}$ correspond to the two inter-cell interference links under consideration. Specifically, $\rho(\psi_{2,1},\theta_{2,1},r_{2,1},\phi_2)$ and
$\rho(\psi_{1,2},\theta_{1,1},r_{1,1},\phi_1)$ are obtained by setting
$(i,k,m)=(2,1,1)$ and $(i,k,m)=(1,2,1)$, respectively.


It is observed from \eqref{Eq:updateR} that obtaining the analytically optimal power allocations $P_{1,1}$ and $P_{2,1}$ for maximizing the sum-rate is challenging due to their coupling in the sum-rate expression  \eqref{Eq:updateR}. A straightforward approach to determine the optimal transmit power allocation is to perform a two-dimensional grid search over the region $[0,P]\times[0,P]$. Moreover, the achievable sum-rate in \eqref{Eq:updateR} can be upper-bounded as
\begin{align}
    R \le   \log_2\left(1+\frac{P_{1,1}N|\beta_{1,1}|^2}{\sigma^2_{1,1}}\right)+\log_2\left(1+\frac{P_{2,1}N|\beta_{2,1}|^2}{\sigma^2_{2,1}}\right),
\end{align}
where the equality holds when both cross-correlation terms vanish, i.e., $\rho(\psi_{2,1},\theta_{2,1},r_{2,1},\phi_2)=\rho(\psi_{1,2},\theta_{1,1},r_{1,1},\phi_1)=0$. This upper bound mainly serves as a performance benchmark to characterize the interference-free limit and to assess the effectiveness of array rotation in suppressing inter-cell mixed-field interference. 

It is noted that the sum-rate in \eqref{Eq:updateR} is significantly influenced by these two cross-correlation terms. Specifically, the sum-rate decreases monotonically as either term increases, and reaches its maximum when both are minimized. The effect of array rotation on the achievable sum-rate is therefore tightly coupled with these terms (mixed-field interference). Consequently, we proceed by examining in detail the influence of array rotation on each term.
Furthermore, the cross-correlation terms are functions of the rotation angles $\phi_1$ and $\phi_2$, with complicated forms given in \eqref{Eq:rho_unified}, making it difficult to theoretically characterize their impact. To address this, we next derive their closed-form approximations based on the Fresnel integrals and analyze the key factors that affect them.
\begin{proposition}\label{The:NF_inter}
    \emph{The \emph{rotation-aware} normalized inter-cell mixed-field interference $\rho(\psi_{i,k},\theta_{i,m},r_{i,m},\phi_i)$ defined in \eqref{Eq:rho_unified} can be approximated as 
   \begin{align}\label{Eq:rho_unified_approx} &\rho\left(\psi_{i,k},\theta_{i,m},r_{i,m},\phi_i\right) \approx G\left(\gamma^{(1)}_{i,k,m},\gamma^{(2)}_{i,m}\right) \nn\\
        &\quad\quad\quad\quad=\left|\frac{\widehat{C}\left(\gamma^{(1)}_{i,k,m},\gamma^{(2)}_{i,m}\right)+j\widehat{S}\left(\gamma^{(1)}_{i,k,m},\gamma^{(2)}_{i,m}\right)}{2\gamma^{(2)}_{i,m}}\right|,
    \end{align}  
     where 
     \begin{align}
        \gamma^{(1)}_{i,k,m} &=(\cos{(\phi_i\!-\!\theta_{i,m})}\!-\!\cos{(\psi_{i,k}\!-\!\phi_i)})\sqrt{\frac{r_{i,m}}{d\sin^2{(\phi_i\!-\!\theta_{i,m})}}},\label{Eq:NF_inter_beta1}\\
        \gamma^{(2)}_{i,m} &=\frac{N}{2} \sqrt{\frac{d\sin^2{(\phi_i-\theta_{i,m})}}{r_{i,m}}}.\label{Eq:NF_inter_beta2}
    \end{align}
    Note that, the general expression in \eqref{Eq:rho_unified_approx} reduces to the specific interference terms $\rho(\psi_{2,1},\theta_{2,1},r_{2,1},\phi_2)$ and $\rho(\psi_{1,2},\theta_{1,1},r_{1,1},\phi_1)$ by setting the indices $(i,k,m)$ to $(2,1,1)$ and $(1,2,1)$, respectively. Here, $\widehat{C}(x,y)=C(x+y)-C(x-y)$ and $\widehat{S}(x,y)=S(x+y)-S(x-y)$, where $C(x)=\int^{x}_{0}\cos(\frac{\pi}{2}t^2){d}t$ and $S(x)=\int^{x}_{0}\sin(\frac{\pi}{2}t^2){d}t$ are the Fresnel integrals.
    }
\end{proposition}
\begin{proof}
     Please refer to Appendix \ref{App0}.
\end{proof}

\begin{figure}[t]
\begin{minipage}{.24\textwidth}
	\centering
\includegraphics[width=1\columnwidth]{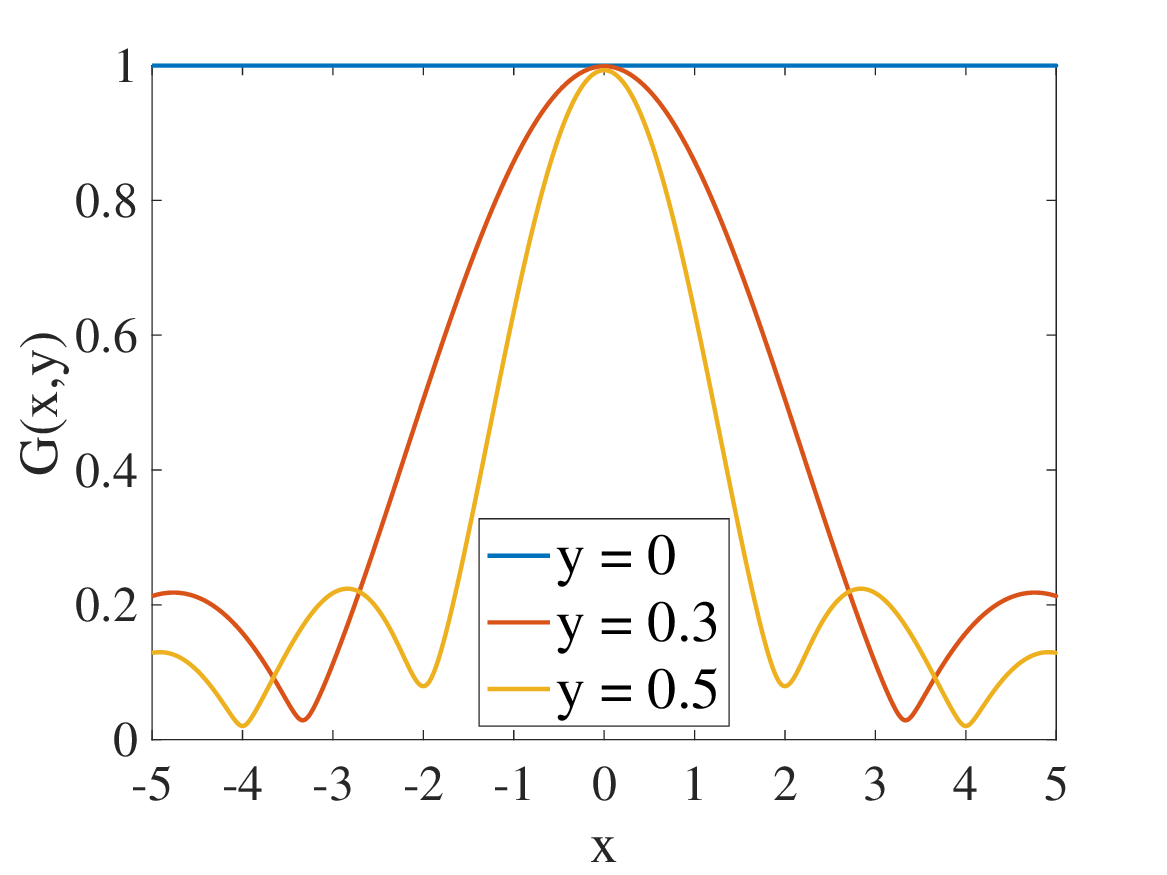}
	\caption{{$G(x,y)$ versus $x$ for fixed $y$.}\label{Fig:corre}} 
    \end{minipage}	
    \hfill
    \begin{minipage}{.24\textwidth}
	\centering
\includegraphics[width=1\columnwidth]{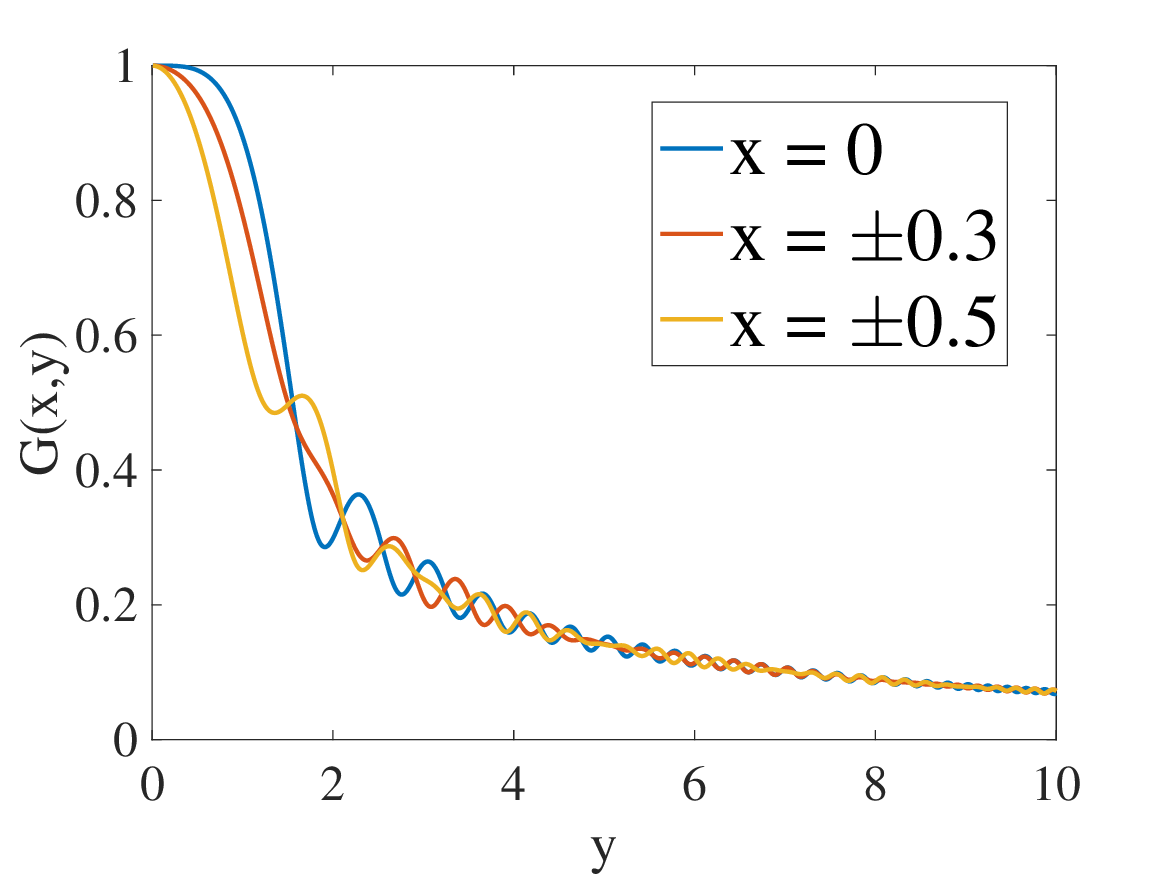}
	\caption{{$G(x,y)$ versus $y$ for fixed $|x|$.}\label{Fig:corre1}} 
    \end{minipage}	
\end{figure}

\begin{remark}(Useful properties of the function $G(x,y)$)
    \emph{We present several key properties that facilitate the analytical characterization of the impact of the two rotation angles on the inter-cell mixed-field interference:
    \begin{itemize}
        \item $G(x,y)$ decreases as $|x|$ increases in either direction, and also decreases as $y$ increases (See Figs. \ref{Fig:corre} and \ref{Fig:corre1}).
        \item The function attains its maximum when either $x=0$ or $y=0$, and reaches its peak value of one when both are zero.
        \item When both rotation angles are set to zero, i.e., $\phi_1=\phi_2=0$, the mixed-field interference expression  simplifies to that of a fixed-antenna configuration, denoted as $G(\tilde{\gamma_1},\tilde{\gamma_2})$ and $G(\hat{\gamma_1},\hat{\gamma_2})$, where:
    \begin{align}
         \tilde{\gamma}^{(1)} &=(\cos{\theta_{2,1}}-\cos{\psi_{2,1}})\sqrt{\frac{r_{2,1}}{d\sin^2{\theta_{2,1}}}},\\
        \tilde{\gamma}^{(2)} &=\frac{N}{2} \sqrt{\frac{d\sin^2{\theta_{2,1}}}{r_{2,1}}},
    \end{align} 
    and
        \begin{align}
         \hat{\gamma}^{(1)} &=(\cos{\theta_{1,1}}-\cos{\psi_{1,2}})\sqrt{\frac{r_{1,1}}{d\sin^2{\theta_{1,1}}}},\\
        \hat{\gamma}^{(2)} &=\frac{N}{2} \sqrt{\frac{d\sin^2{\theta_{1,1}}}{r_{1,1}}}.
    \end{align} 
    \end{itemize}
    }
\end{remark}
\begin{remark}(Key factors affecting the mixed-field interference)
    \emph{The mixed-field interference in single-cell and multi-cell scenarios is influenced by different key factors \cite{zhangRAmixed}.
    To be specific, in the single-cell setting, the \emph{intra-cell} mixed-field interference between far-field and near-field users is primarily determined by the intra-cell angle and range of the near-field user, as well as the intra-cell angle of the far-field user. In contrast, in the multi-cell case, the \emph{inter-cell} mixed-field interference experienced by ${\rm U}_{1,1}$ in cell $1$ from ${\rm U}_{2,1}$ in cell $2$, denoted as $\rho(\psi_{2,1},\theta_{2,1},r_{2,1},\phi_2)$ and characterized by \eqref{Eq:rho_unified} and \eqref{Eq:rho_unified_approx}, depends on the intra-cell angle and range of ${\rm U}_{2,1}$ in cell $2$, as well as the \emph{inter-cell angle} of ${\rm U}_{1,1}$ relative to ${\rm BS}_2$. This highlights a fundamental distinction: while single-cell mixed-field interference is directly determined by the intra-cell angles of the users, inter-cell mixed-field interference is governed by the inter-cell angle rather than the intra-cell angle. Furthermore, the intensity of the inter-cell interference is strongly influenced by the intra-cell angle and range of users in adjacent cells.
    This difference reflects the more complex propagation geometry inherent in multi-cell environments.
    }
\end{remark}

Next, we characterize the impact of rotation angles $\phi_1$ and $\phi_2$ on 
the inter-cell mixed-field interference terms, namely, $\rho(\psi_{2,1},\theta_{2,1},r_{2,1},\phi_2)$ and $\rho(\psi_{1,2},\theta_{1,1},r_{1,1},\phi_1)$.
From \eqref{Eq:rho_unified}, it is evident that these interference terms are strongly influenced by their respective rotation angles, $\phi_1$ and $\phi_2$. This dependence introduces a crucial new design degree-of-freedom (DoF). It not only enables the analytical characterization of how rotation angles affect the inter-cell mixed-field interference, but also offers an effective means to mitigate such interference by strategically adjusting these parameters. 
In particular, the interference term $\rho(\psi_{2,1},\theta_{2,1},r_{2,1},\phi_2)$, representing the inter-cell mixed-field interference from ${\rm U}_{2,1}$ in the neighboring ${\rm BS}_2$ to ${\rm U}_{1,1}$, depends solely on the rotation angle $\phi_2$ of ${\rm BS}_2$ and is unaffected by the rotation angle $\phi_1$ of its serving ${\rm BS}_1$. Consequently, for ${\rm U}_{1,1}$, the interference caused by ${\rm U}_{2,1}$ can be reduced by properly designing $\phi_2$, whereas adjusting $\phi_1$ has no influence on this interference. Furthermore, the geometric dependence between ${\rm U}_{1,1}$ and ${\rm U}_{2,1}$ arises through the \emph{inter-cell angle} of ${\rm U}_{1,1}$ with respect to ${\rm BS}_2$ (i.e., $\psi_{2,1}$).
This observation allows for a decoupled analysis of rotation-angle effects on the corresponding inter-cell mixed-field interference. Accordingly, we focus on the impact of the rotation angle $\phi_2$ on $\rho(\psi_{2,1},\theta_{2,1},r_{2,1},\phi_2)$, as the other interference term, $\rho(\psi_{1,2},\theta_{1,1},r_{1,1},\phi_1)$, follows the same analytical framework.
To theoretically demonstrate the interference suppression benefits of $\phi_2$, we formulate the problem as follows:

\begin{subequations}
	\begin{align}
		({\bf P2}):\min_{\substack{\phi_2 }}  ~~& \rho(\psi_{2,1},\theta_{2,1},r_{2,1},\phi_2)
		\\
		\text{s.t.}~~  &\phi_2 \in \left[\phi_{2,{\rm \min}},\phi_{2,{\rm \max}}\right].
	\end{align}
 \end{subequations}
 
The optimal solution to problem (P2) is generally not available in closed form due to the complicated dependence of $\rho(\psi_{2,1},\theta_{2,1},r_{2,1},\phi_2)$ on the rotation angle. A straightforward approach to solving problem (P2) is to perform a one-dimensional exhaustive search over the admissible rotation angle range via quantization. While this method can yield a high-quality solution, it offers few insights into the impact of array rotation on the multi-cell mixed-field interference. To address this issue, we next present several key theoretical insights that uncover the fundamental characteristics of array rotation in interference suppression, derived from the solution to problem (P2) from the perspectives of angle and range.

\begin{corollary}(Effect of user angle)\label{Pro:equal}
    \emph{The effect of the intra-cell angle of ${\rm U}_{2,1}$ ($\theta_{2,1}$) on the normalized inter-cell mixed-field interference $\rho(\psi_{2,1},\theta_{2,1},r_{2,1},\phi_2)$ depends heavily on the inter-cell angle of ${\rm U}_{1,1}$ ($\psi_{2,1}$) and can be categorized into the following three cases:
    \begin{itemize}
        \item \textbf{Case 1:} $(\psi_{2,1}=\theta_{2,1}=\frac{\pi}{2})$: Both ${\rm U}_{1,1}$ and ${\rm U}_{2,1}$ are located in the boresight direction. The suboptimal solution to problem (P2) is $\phi_2^*=0$,  indicating that the array rotation does not provide any benefit for suppressing inter-cell mixed-field interference.
        \item \textbf{Case 2:} $(\psi_{2,1}=\theta_{2,1}\neq \frac{\pi}
        {2})$: In this case, the suboptimal solution to problem (P2) is 
        \begin{align}
            \phi_2^* = 
            \begin{cases}
                \min(\psi_{2,1}-\frac{\pi}{2},\phi_{\rm max}),~\text{if}~\psi_{2,1} > \frac{\pi}{2}, \\
                \max(\psi_{2,1}-\frac{\pi}{2},\phi_{\rm min}),~\text{if}~\psi_{2,1} < \frac{\pi}{2},
            \end{cases}
        \end{align}
        implying that array rotation can effectively mitigate the interference.
        \item \textbf{Case 3:} $(\psi_{2,1} \neq \theta_{2,1})$: There always exists a rotation angle $\phi_2$ such that $\gamma_1>\tilde{\gamma_1}$ and $\gamma_2>\tilde{\gamma_2}$, and hence 
        \begin{equation}
    \rho(\psi_{2,1},\theta_{2,1},r_{2,1},\phi_2)<\rho(\psi_{2,1},\theta_{2,1},r_{2,1},0),
        \end{equation}
        thereby suppressing the interference.
    \end{itemize}
    }
\end{corollary}
\begin{proof}
Please refer to Appendix \ref{App2}.
\end{proof}

\begin{corollary}(Effect of user range)\label{Pro:userrange}
    \emph{When the inter-cell angle of ${\rm U}_{1,1}$ equals the intra-cell angle of ${\rm U}_{2,1}$, i.e., $\psi_{2,1}=\theta_{2,1}$, the inter-cell mixed-field interference experienced by ${\rm U}_{1,1}$ from  ${\rm U}_{2,1}$ generally increases as the range $r_{2,1}$ of ${\rm U}_{2,1}$ grows. Moreover, the benefits of array rotation in mitigating this interference become marginal, especially at greater distances.}
\end{corollary}
\begin{proof}
 Please refer to Appendix \ref{App3}.
\end{proof}
Corollary \ref{Pro:userrange} is intuitively expected since when the inter-cell angle of ${\rm U}_{1,1}$ is equal to the intra-cell angle of ${\rm U}_{2,1}$, the increasing range of ${\rm U}_{2,1}$ will inevitably weaken its near-field effect, causing it to exhibit the far-field characteristics. Consequently, ${\rm U}_{2,1}$ effectively behaves as a far-field user, leading to increased inter-cell mixed-field interference experienced by ${\rm U}_{1,1}$  \cite{zhang2023mixed}. 

    \begin{figure*}[t!]
\begin{minipage}{.328\textwidth}
	\centering
\includegraphics[width=1\columnwidth]{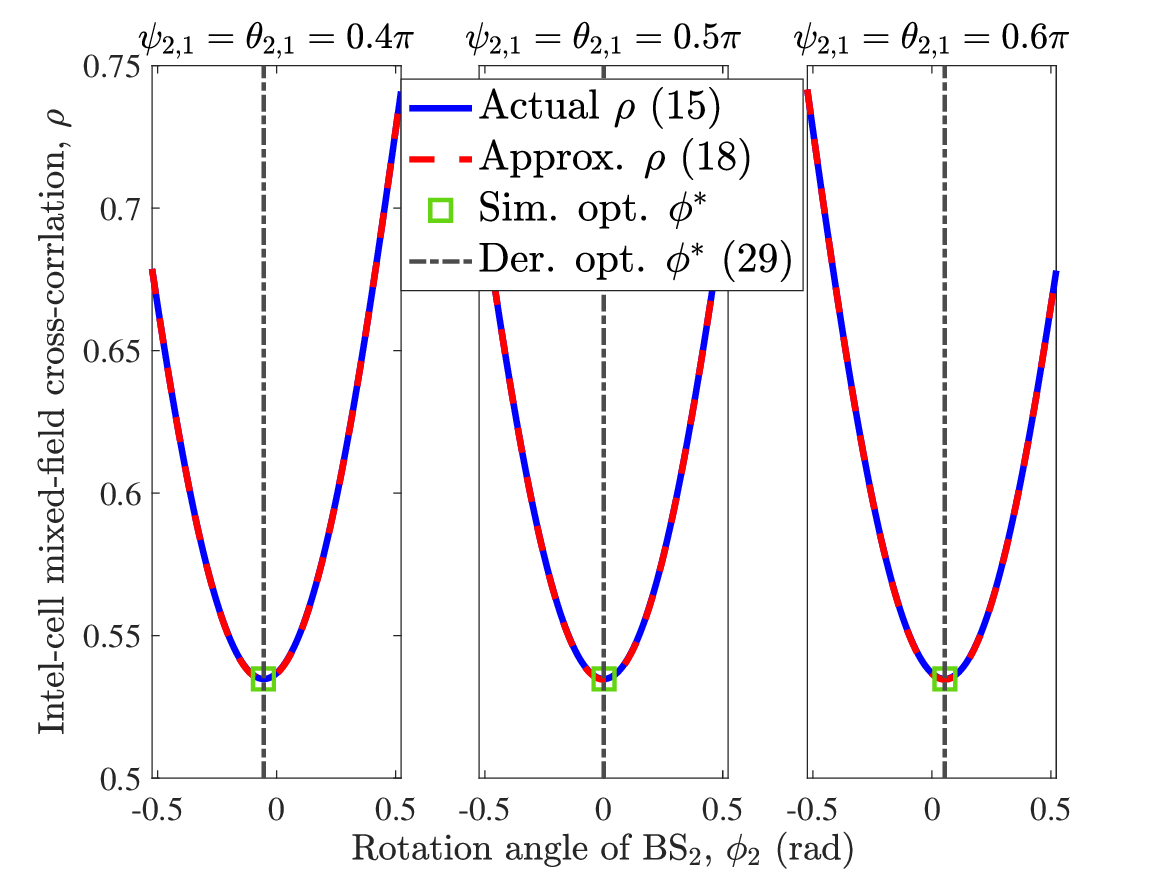}
	\caption{{Inter-cell mixed-field interference power versus rotation angle.}\label{Fig:verify}} 
    \end{minipage}	
    \hfill
    \begin{minipage}{.328\textwidth}
	\centering
\includegraphics[width=1\columnwidth]{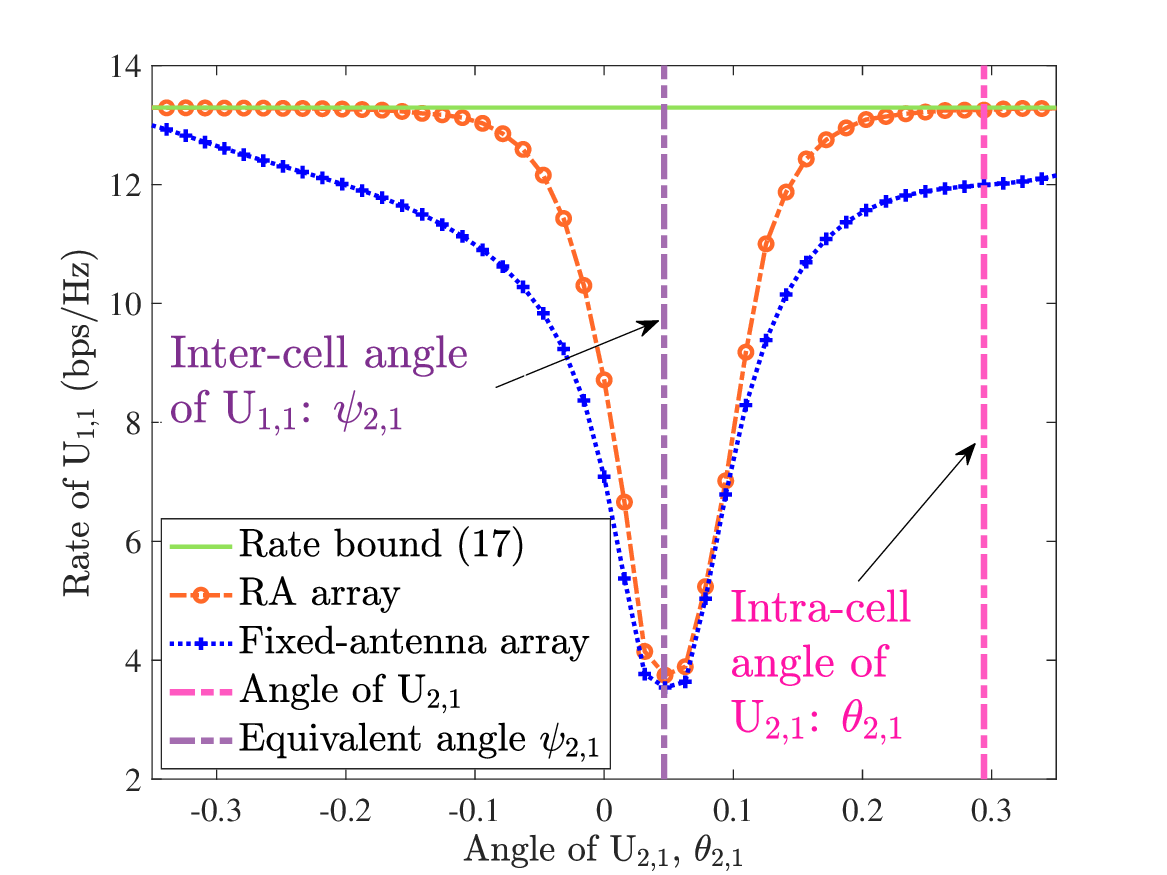}
	\caption{{Rate of ${\rm U}_{1,1}$ versus angle of ${\rm U}_{2,1}$ at $r_{2,1} = 0.3Z_{\rm Ray}$.}\label{Fig:angle}} 
    \end{minipage}	
    \hfill
        \begin{minipage}{.328\textwidth}
\centering
\includegraphics[width=1\columnwidth]{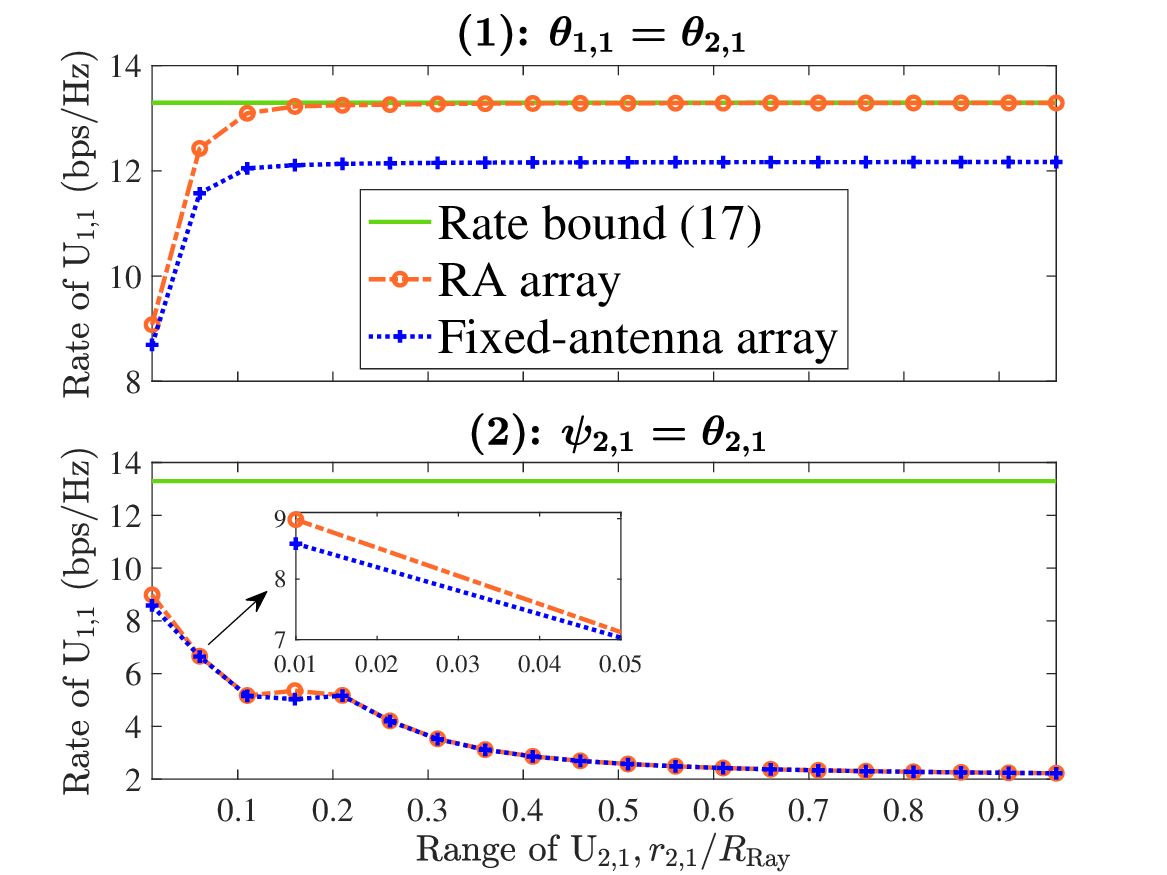}
	\caption{{Rate of ${\rm U}_{1,1}$ versus range of ${\rm U}_{2,1}$ at $\theta_{1,2} = \theta_{2,1}= 0.4\pi$ and  $\psi_{2,1} = \theta_{2,1}= 0.4\pi$.}\label{Fig:dist}} 
    \end{minipage}	
\end{figure*}
\begin{example}(How does the rotation angle affect inter-cell mixed-field interference?)
    \emph{To demonstrate the advantages of RAs in mitigating complex inter-cell mixed-field interference, we present the following example. Specifically, we first verify the accuracy of the closed-form approximation for the inter-cell mixed-field interference derived in Lemma \ref{The:NF_inter}. As shown in Fig. \ref{Fig:verify}, the approximation in \eqref{Eq:rho_unified_approx} closely matches the exact interference expression in \eqref{Eq:rho_unified}. Moreover, the optimal rotation angles derived in Corollary \ref{Pro:equal} align well with the numerically obtained values across a variety of spatial-angle configurations. Next, we investigate the effect of rotation angle on the achievable rate of ${\rm U}_{1,1}$ with respect to the intra-cell angle and range of ${\rm U}_{2,1}$. In particular, Figs. \ref{Fig:angle} and \ref{Fig:dist} plot the rate of ${\rm U}_{1,1}$ versus the intra-cell angle and  range of ${\rm U}_{2,1}$, respectively, under $N=129$ and $f=28$ GHz. The main observations are summarized below. 
    \begin{itemize}
        \item \emph{Effect of intra-cell angle of ${\rm U}_{2,1}$:} An interesting observation from Fig. \ref{Fig:angle} is that the lowest rate of ${\rm U}_{1,1}$ occurs when the intra-cell angle of ${\rm U}_{2,1}$ coincides with the \emph{inter-cell angle} of ${\rm U}_{1,1}$ relative to ${\rm BS}_2$, i.e., $\psi_{2,1}$, rather than with the intra-cell angle of ${\rm U}_{1,1}$. This aligns well with Corollary \ref{Pro:equal} and is significantly different from the conventional single-cell mixed-field interference, where the lowest rate typically occurs when the two users share the same intra-cell angle \cite{zhang2023mixed}. 
        \item \emph{Effect of range of ${\rm U}_{2,1}$:} To comprehensively illustrate the impact of the range of ${\rm U}_{2,1}$, we consider two cases: (1) $\theta_{1,1} = \theta_{2,1}$, i.e., the intra-cell angles of ${\rm U}_{1,1}$ and ${\rm U}_{2,1}$ are the same; and (2) $\psi_{2,1}=\theta_{2,1}$, i.e., the inter-cell angle of ${\rm U}_{1,1}$ equals the intra-cell angle of ${\rm U}_{2,1}$. In the first case, the rate of ${\rm U}_{1,1}$ attained by both RA and fixed-antenna arrays increases as the range of ${\rm U}_{2,1}$ grows. This is because, although the intra-cell angles are equal, the mixed-field interference is primarily determined by the difference between the inter-cell angle of ${\rm U}_{1,1}$ and the intra-cell angle ${\rm U}_{2,1}$. As the range increases, the interference approaches the far-field regime, where the angular separation effectively suppresses interference. In the second case, the mixed-field interference strongly depends on the range of ${\rm U}_{2,1}$, and generally increases with the range of ${\rm U}_{2,1}$, thereby resulting in a declining rate. Furthermore, in this case, the interference-mitigation capability of the rotation angle becomes very limited as the range of ${\rm U}_{2,1}$ increases. These observations are consistent with Corollary \ref{Pro:userrange}. 
    \end{itemize}
    }
\end{example}
 \begin{figure}[t]
\begin{minipage}{.241\textwidth}
	\centering
\includegraphics[width=1\columnwidth]{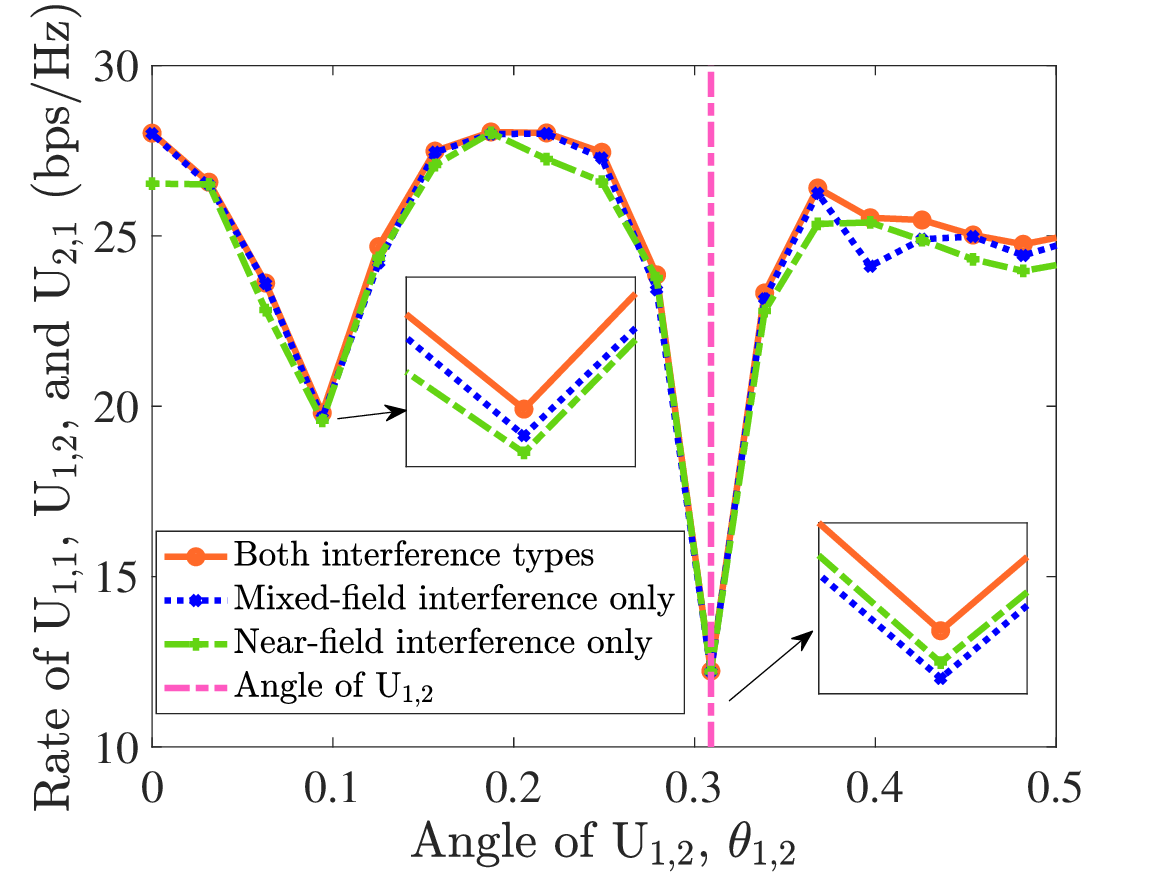}
	\caption{{Effect of angle of ${\rm U}_{1,2}$.}\label{Fig:sc3angle}} 
    \end{minipage}	
    \hfill
    \begin{minipage}{.241\textwidth}
	\centering
\includegraphics[width=1\columnwidth]{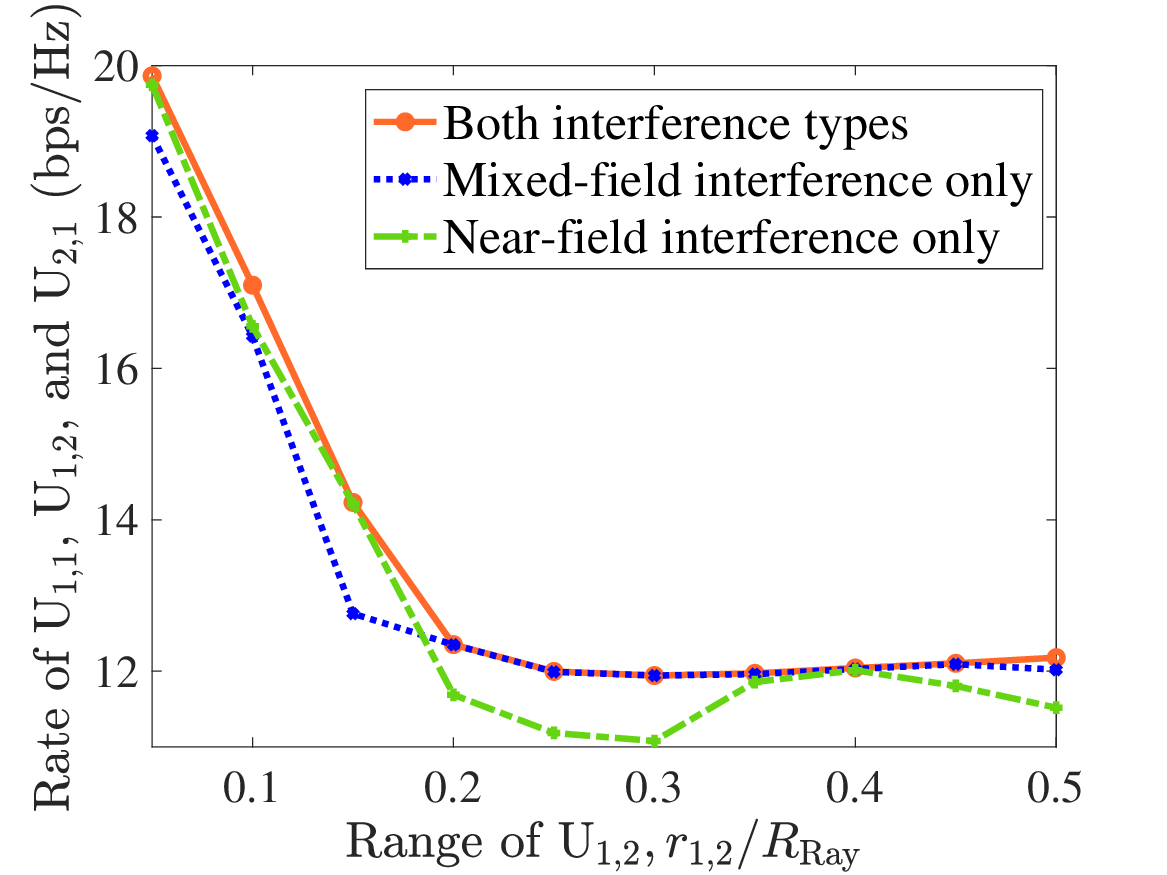}
	\caption{{Effect of range of ${\rm U}_{1,2}$.}\label{Fig:sc3range}} 
    \end{minipage}	
\end{figure}

\subsection{Trade-Off in Array Rotation for Mitigating Near-Field and Mixed-Field Interference}\label{Sec:tradoff}
It is worth emphasizing that, in practice, the achievable rate of a multi-cell mixed-field system is jointly determined by \emph{intra-cell near-field interference} and \emph{inter-cell mixed-field interference}. Consequently, the RA array rotation at each BS must be carefully designed to suppress both types of interference. However, this design naturally encounters a fundamental \emph{trade-off}, as the two types of interference are tightly coupled in the achievable sum-rate yet exhibit distinct spatial patterns, thereby complicating theoretical analysis. To clearly illustrate this trade-off, we extend the special case in Section \ref{Sec:ana} by introducing an additional user in cell $1$, denoted as ${\rm U}_{1,2}$. 
Specifically, in Figs. \ref{Fig:sc3angle} and \ref{Fig:sc3range}, we plot the achievable sum-rate of ${\rm U}_{1,1}$, ${\rm U}_{1,2}$, and ${\rm U}_{2,1}$ under varying angle and range of ${\rm U}_{1,2}$ while keeping ${\rm U}_{1,1}$ and ${\rm U}_{2,1}$ fixed at $\theta_{1,1}=\theta_{2,1}=0.4\pi$ and $r_{1,1}=r_{2,1}=0.3Z_{\rm Ray}$. For comparison, we consider two benchmark schemes: \emph{1) Near-field interference only:} the array rotation is optimized to mitigate intra-cell near-field interference exclusively; \emph{2) Mixed-field interference only:} the array rotation is optimized to mitigate inter-cell mixed-field interference exclusively. From Fig. \ref{Fig:sc3angle}, we observe that the most significant rate loss caused by ignoring near-field interference and mixed-field interference occurs when the intra-cell angle of ${\rm U}_{1,2}$ matches the intra-cell angle of ${\rm U}_{1,1}$ and the inter-cell angle, respectively. Moreover, the ``mixed-field interference only” scheme achieves superior performance, since variations in the angle of ${\rm U}_{1,2}$ have a stronger impact on near-field interference. In Fig. \ref{Fig:sc3range}, we further observe that when the range of ${\rm U}_{1,2}$ is small (e.g., $r_{1,2}/R_{\rm Ray}\le 0.18$), near-field interference dominates, and ignoring it results in severe performance loss. In contrast, as the range of ${\rm U}_{1,2}$ increases, mixed-field interference becomes more prominent, and neglecting it leads to even larger performance degradation. 

These results fundamentally reveal the trade-off in RA array rotation for mitigating near-field and mixed-field interference. Intuitively, the rotation configuration should balance the two types of interference depending on their relative strengths. In particular, when intra-cell near-field interference is moderate, the array rotation should prioritize suppressing inter-cell mixed-field interference, whereas when the inter-cell mixed-field interference is moderate, the array rotation should instead emphasize suppressing intra-cell near-field interference.

\begin{remark}[Practical implementation issues of RA-enabled multi-cell mixed-field systems]
    \emph{The rotation speed and response time of RA play an important role in enabling effective real-time interference suppression. If these parameters are not properly designed, the performance of the proposed scheme may be significantly degraded. In practice, RA can be implemented using either mechanical or electronic rotation mechanisms \cite{zheng2025rotatableM}. Mechanically driven RA, particularly those based on micro-electromechanical system (MEMS) actuators, typically exhibit response times ranging from microseconds to milliseconds. In contrast, electronically driven RA can achieve response times from nanoseconds to milliseconds, which enables adaptation to both slowly varying and rapidly fluctuating channel conditions \cite{zheng2025rotatableM}. Moreover, practical implementations of rotatable antennas, including both mechanical and electronic realizations, can achieve angular positioning accuracies on the order of hundredths to thousandths of the signal wavelength \cite{11197972}. To further account for practical latency constraints, a two-timescale optimization framework can be adopted, where RA rotation angles are optimized using long-term statistical CSI over a large timescale, while short-term beamforming is adapted based on instantaneous CSI. Under this framework, the proposed instantaneous-CSI-based design serves as a performance upper bound and provides valuable insights into the achievable gains of RA-enabled multi-cell mixed-field communication systems.}
\end{remark}

\section{Proposed Algorithm to Problem (P1)}\label{Sec:Proposed}
In this section, we propose an efficient algorithm to obtain a high-quality solution to problem (P1) by equivalently decomposing it into two subproblems: an inner layer subproblem for optimizing the hybrid beamforming matrices at each BS for given rotation angles, and an outer layer subproblem that determines the rotation angles of the RA arrays. The formulation and solution of these subproblems are presented in detail below.

\subsection{Inner Layer: Beamformer Optimization} \label{Sec:BF_opt}
Given any feasible rotation angles of all RA arrays $\boldsymbol{\phi}$, the inner-layer problem of beamformer optimization is formulated as follows:
\begin{subequations}
	\begin{align}
		({\bf P3}):~\max_{\substack{\{\mathbf{F}_{{\rm A},m}, \mathbf{F}_{{\rm D},m}\}}}  ~&\sum^{M}_{m=1}\sum^{K}_{k=1} R_{m,k}\left(\left\{\mathbf{F}_{{\rm A},m},\mathbf{F}_{{\rm D},m}\right\}\right)\label{P4:obj}
		\\
		\text{s.t.}~~
		&	\|\mathbf{F}_{{\rm A},m}\mathbf{F}_{{\rm D},m}\|^2_F\le P,\forall{m},\label{P4:pow_cons}\\
        & |\mathbf{F}_{{\rm A},m}(i,j)|=1,\forall{(i,j)}\in\mathcal{F}.\label{P4:ana_norm}
	\end{align}
 \end{subequations}
Note that problem (P3) remains challenging to solve directly due to the non-convex objective function in \eqref{P4:obj} and the unit-modulus constraint in \eqref{P4:ana_norm}. Moreover, the coupling of the hybrid beamforming matrices in both the objective function and constraints further complicates the problem.
To tackle these challenges, we adopt an efficient two-stage beamforming framework to obtain a high-quality solution to problem (P3) \cite{liu2023near}.

\subsubsection{Stage 1 (Analog beamformer design)} In the first stage, we design the analog beamforming matrix $\mathbf{F}_{{\rm A},m}$ at the BS in cell $m$ to maximize the received signal power at its associated near-field users. Specifically, for each user $k$, the corresponding column of $\mathbf{F}_{{\rm A},m}$, denoted by $\mathbf{f}_{{\rm A},m,k}$, is constructed as
\begin{equation}
    \mathbf{f}_{{\rm A},m,k} = \sqrt{N}\mathbf{b}(\theta_{m,k},r_{m,k},\phi_m),~\forall{k\in \mathcal{K}_{m}}.
\end{equation}

\subsubsection{Stage 2 (Digital beamformer design)}
For the BS in cell $m$, given $\mathbf{F}_{{\rm A},m}$, the optimization of the digital beamforming matrix proceeds as follows. Specifically, based on the obtained analog beamformer $\mathbf{F}_{{\rm A},m}$, the effective channels for the near-field users in cell $m$ are computed as:
\begin{align}
    \Bar{\mathbf{h}}_{m,m,k} &= \mathbf{F}^H_{{\rm A},m}{\mathbf{h}}_{m,m,k},~\forall{m\in\mathcal{M}},\forall{k\in\mathcal{K}_m}, \\
     \Bar{\mathbf{h}}_{i,m,k} &= \mathbf{F}^H_{{\rm A},i}{\mathbf{h}}_{i,m,k},~\forall{i\in\mathcal{M}\setminus\{m\}},\forall{k\in\mathcal{K}_m}.
\end{align}
Accordingly, the achievable rate at user $k$ in cell $m$, denoted by $R_{m,k}$ in \eqref{Eq:userrate}, can be equivalently expressed as shown in \eqref{Eq:userrate1} at the bottom of the next page. Subsequently, we develop an efficient approach for determining the digital beamforming matrix at each BS by leveraging the SDR and SCA techniques. To this end, we define the following auxiliary variables: 
$\mathbf{H}_{m,m,k} = \Bar{\mathbf{h}}_{m,m,k}\Bar{\mathbf{h}}_{m,m,k}^H$, $\mathbf{H}_{i,m,k} = \Bar{\mathbf{h}}_{i,m,k}\Bar{\mathbf{h}}_{i,m,k}^H$ and $\mathbf{W}_{m,k} = \mathbf{f}_{{\rm D},m,k}\mathbf{f}_{{\rm D},m,k}^H$. Problem (P3) can then be equivalently reformulated as
\begin{figure*}[b]
\hrulefill
    \begin{align}\label{Eq:userrate1}
    R_{m,k} = \log_2\left(1+\frac{|\Bar{\mathbf{h}}^{H}_{m,m,k}\mathbf{f}_{{\rm D},m,k}|^2}{\sum^{K}_{\ell=1,\ell\neq k}|\Bar{\mathbf{h}}^{H}_{m,m,k}\mathbf{f}_{{\rm D},m,\ell}|^2+\sum^{M}_{i=1,i\neq m}\sum^{K}_{j=1}|\Bar{\mathbf{h}}^{H}_{i,m,k}\mathbf{f}_{{\rm D},i,j}|^2+\sigma^2_{m,k}}\right)
\end{align}
\end{figure*}
\begin{subequations}
	\begin{align}
		({\bf P4}):~\min_{\substack{\{\mathbf{W}_{m,k}\}}}  ~&-\sum^{M}_{m=1}\sum^{K}_{k=1} \tilde{R}_{m,k}\label{P5:obj}
		\\
		\text{s.t.}~~
		&\eqref{P4:pow_cons},\label{P5:pow}\\
        & \mathbf{W}_{m,k}\succeq \mathbf{0},~\forall{m,k}\label{P5:sem}\\
         & \operatorname{Rank}(\mathbf{W}_{m,k})\le1,~\forall{m,k},\label{P5:rank}
	\end{align}
 \end{subequations}
where $\tilde{R}_{m,k}$ is shown in \eqref{Eq:newR} at the bottom of the next page. Constraints \eqref{P5:sem} and \eqref{P5:rank} are imposed to ensure $\mathbf{W}_{m,k} = \mathbf{f}_{{\rm D},m,k}\mathbf{f}_{{\rm D},m,k}^H$ upon optimization. Moreover, to address the non-convexity of the objective function in problem (P4), we first rewrite it as a difference of convex functions, i.e., $-\sum^{M}_{m=1}\sum^{K}_{k=1} \tilde{R}_{m,k} = N_1-D_1$, where 
\begin{figure*}[b]
\hrulefill
    \begin{align}\label{Eq:newR}
    \tilde{R}_{m,k} = \log_2\left(1+\frac{\operatorname{Tr}(\mathbf{H}^{H}_{m,m,k}\mathbf{W}_{m,k})}{\sum^{K}_{\ell=1,\ell\neq k}\operatorname{Tr}(\mathbf{H}^{H}_{m,m,k}\mathbf{W}_{m,\ell})+\sum^{M}_{i=1,i\neq m}\sum^{K}_{j=1}\operatorname{Tr}(\mathbf{H}^{H}_{i,m,k}\mathbf{W}_{i,j})+\sigma^2_{m,k}}\right)
\end{align}
\end{figure*}
\begin{align}\label{Eq:P3_Coe}
        N_{1}  &=-\sum^{M}_{m=1}\sum^{K}_{k=1}\log_2\bigg(\sum^{K}_{i=1}\mathrm{Tr}(\mathbf{H}_{m,m,k}\mathbf{W}_{m,k})\nn\\
&\quad\quad\quad+\sum^{M}_{i=1,i\neq m}\sum^{K}_{j=1}\operatorname{Tr}(\mathbf{H}^{H}_{i,m,k}\mathbf{W}_{i,j})+\sigma^2_{m,k}\bigg), \\
        D_{1} & = -\sum^{M}_{m=1}\sum^K_{k=1}\log_2\bigg(\sum^{K}_{\ell=1,\ell\neq k}\operatorname{Tr}(\mathbf{H}^{H}_{m,m,k}\mathbf{W}_{m,\ell})\nn\\
         &\quad\quad\quad+\sum^{M}_{i=1,i\neq m}\sum^{K}_{j=1}\operatorname{Tr}(\mathbf{H}^{H}_{i,m,k}\mathbf{W}_{i,j})+\sigma^2_{m,k}\bigg).
\end{align}
Next, we adopt the SCA method to iteratively obtain a convex upper bound for the objective function. Specifically, we construct a global underestimator for $D_1$, where we adopt the superscript $(t)$ to denote the iteration index of the optimization variables. For any feasible point ${\mathbf{W}^{(t)}}=\left\{\mathbf{W}^{(t)}_{m,k}\right\}$, $\forall{m,k}$, a lower bound of $D_1$ can be obtained via its first-order Taylor approximation, expressed as:
\begin{align}
    D_{1}&\left(\mathbf{W}\right)  \ge D_1\left(\mathbf{W}^{(t)}\right)\nn\\
    &+ \operatorname{Tr}\left(\nabla^H_{\mathbf{W}_{m,k}}D_{1}\left(\mathbf{W}^{(t)}_{m,k}\right)\left(\mathbf{W}_{m,k}-\mathbf{W}^{(t)}_{m,k}\right)\right)\nn\\
     &\triangleq \tilde{D}_{1}\left(\mathbf{W},\mathbf{W}^{(t)}\right),
\end{align}
where the gradient of $D_1$ with respect to $\mathbf{W}_{m,k}$ is given by
    \begin{align}\label{Eq:gradient}
        \nabla_{\mathbf{W}_{m,k}} D_{1}(\mathbf{W}_{m,k})=&-\frac{1}{\ln 2}\bigg( \sum_{k'=1, k' \neq k}^{K} \frac{\mathbf{H}_{m,m,k'}}{a_{m,k'}} \nn\\
        &+ \sum_{m'=1, m' \neq m}^{M} \sum_{k'=1}^{K} \frac{\mathbf{H}_{m,m',k'}}{a_{m',k'}} \bigg),
    \end{align}
with
\begin{align}
     a_{m,k} = &\sum_{\ell=1, \ell \neq k}^{K} \operatorname{Tr}(\mathbf{H}_{m,m,k}^{H} \mathbf{W}_{m,\ell}) \nn\\
     &+ \sum_{i=1, i \neq m}^{M} \sum_{j=1}^{K} \operatorname{Tr}(\mathbf{H}_{i,m,k}^{H} \mathbf{W}_{i,j}) + \sigma_{m,k}^2.
\end{align}
By substituting the convex upper bound $\tilde{D}_{1}\left(\mathbf{W},\mathbf{W}^{(t)}\right)$ into the objective function in \eqref{P5:obj}, the only remaining non-convexity in problem (P4) arises from the rank constraint \eqref{P5:rank}. To address this issue, we employ SDR by dropping the rank constraint \eqref{P5:rank}. Therefore, the resultant relaxed problem to be solved at feasible point ${\mathbf{W}^{(t)}}$ is given by
\begin{subequations}
	\begin{align}
		({\bf P5}):~\min_{\substack{\{\mathbf{W}_{m,k}\}}}  ~&N_1-\tilde{D}_{1}\label{P6:obj}
		\\
		\text{s.t.}~~
		&\eqref{P5:pow},\eqref{P5:sem},
	\end{align}
 \end{subequations}
The relaxed problem (P5) is convex with respect to the optimization variables and can thus be effectively tackled using standard solvers, such as CVX. More importantly, the tightness of the SDR method for such problems has been extensively validated in the literature, ensuring that a rank-one optimal solution to problem (P5) can always be obtained.  This procedure is well established and therefore omitted here for brevity. For further information, the interested reader is referred to \cite{9913311}.

\subsection{Outer Layer: Rotation Angle Optimization}\label{Sec:Rot_Opt}
With the digital beamformers optimized at all BSs, the outer-layer subproblem focuses on optimizing the rotation angles of the RA arrays. This subproblem is formulated as follows:\footnote{Notice that the evaluation of rotation-dependent channels for different rotation angles in the PSO-based outer-layer search relies heavily on parametric channel estimation, which is generally more challenging to obtain than conventional pilot-based channel estimation schemes (see, e.g., \cite{9940281,zhang2022fast,wu2023near,9693928}). A practical approach to acquiring such information is to adopt an iterative framework consisting of two main procedures, namely CSI estimation and RA orientation adjustment, which are carried out alternately during each channel training period \cite{11134688}.}
\begin{subequations}
	\begin{align}
		({\bf P6}):~\max_{\substack{\boldsymbol{\phi} }}  ~&\sum^{M}_{m=1}\sum^{K}_{k=1} R_{m,k}\left(\left\{\mathbf{F}_{{\rm A},m},\mathbf{F}_{{\rm D},m}\right\},\boldsymbol{\phi}\right)
		\\
		\text{s.t.}~~
		&	\eqref{P1:rot_angle}.\nn
	\end{align}
 \end{subequations}
For this outer-layer problem, a closed-form solution for the optimized rotation angles is generally unattainable, as it depends on solving the inner problem (P5). Furthermore, the complex coupling between the far-field and near-field steering vectors with respect to the rotation angle vector $\boldsymbol{\phi}$ increases the problem’s complexity. Although existing methods, such as the quasi-Newton method \cite{rotationISAC}, can be employed to optimize $\boldsymbol{\phi}$, they are prone to converge to a local
optimum or may yield a low-quality solution. To address this issue, we adopt a PSO algorithm to obtain a high-quality solution to problem (P6).
Specifically, the steps of the PSO algorithm for optimizing the rotation angles of the RA arrays are described below.

\subsubsection{{Initialization}} First, to perform the PSO-based method, a population of $S$ feasible rotation angle vectors for the RA arrays is randomly generated within the predefined solution space, which is given by $\mathcal{S}^{(0)}=\left\{\boldsymbol{\phi}^{(0)}_{1},\ldots,\boldsymbol{\phi}^{(0)}_{s},\ldots,\boldsymbol{\phi}^{(0)}_{S}\right\}.$
Each particle $s\in\mathcal{S}$ represents a feasible configuration of $M$ rotation angles, given by $\boldsymbol{\phi}^{(0)}_{s} = \left[ {\phi}^{(0)}_{s,1}, {\phi}^{(0)}_{s,2},\dots, {\phi}^{(0)}_{s,M}\right]^T.$
This applies to the $s$-th rotation angle vector in the $t$-th PSO iteration, where $t\in\mathcal{T}=\{1,2,\ldots,T\}$, and $T$ is the maximum iteration number for the PSO-based method.

\subsubsection{PSO operators} Next, we update the particles in each PSO iteration by conducting operations on their position and associated velocity. Specifically, for each particle $s$, the initial velocity is defined as:
\begin{equation}
    \mathbf{v}^{(0)}_{s} = \left[{v}^{(0)}_{s,1}, {v}^{(0)}_{s,2},\dots, {v}^{(0)}_{s,M}\right]^T.
\end{equation}
At each iteration $t$, the velocity and position of each particle are updated as: 
\begin{align}
    \mathbf{v}^{(t+1)}_{s} &\!=\! \omega\mathbf{v}^{(t)}_{s}\!+\!c_1\tau_1(\boldsymbol{\phi}_{s,{\rm p}}\!-\!\boldsymbol{\phi}^{(t)}_{s})\!+\!c_2\tau_2(\boldsymbol{\phi}_{{\rm g}}-\boldsymbol{\phi}^{(t)}_{s}),\\
     \boldsymbol{\phi}^{(t+1)}_{s}&= \boldsymbol{\phi}^{(t)}_{s}+ \mathbf{v}^{(t+1)}_{s},
\end{align}
where $\boldsymbol{\phi}_{s,{\rm p}}$ and $\boldsymbol{\phi}_{{\rm g}}$ represent the individual best position of the $s$-th particle
and the global best position across all particles, respectively. The parameter $\omega \in [\omega_{\rm min},\omega_{\rm max}]$ is the inertia weight, while $c_1$ and $c_2$ denote the individual and global learning factors, indicating the step sizes toward the individual and global best positions, respectively. The scalars $\tau_1,\tau_2\in
[0,1]$ are uniformly distributed random variables that introduce stochasticity into the search process. In addition, to ensure that the rotation angles comply with the constraint \eqref{P1:rot_angle} of problem (P1), which requires each rotation angle $\phi_m$ to lie within a predefined admissible range, 
any rotation angle that violates this constraint
 is projected back onto the feasible interval as:
\begin{align}
    [\boldsymbol{\phi}^{(t)}_{s}]_m = \max\{\min\{[\boldsymbol{\phi}^{(t)}_{s}]_m,\phi_{m,{\rm max}}\},\phi_{m,{\rm min}}\}.
\end{align}

\subsubsection{Fitness function} The fitness (or quality) of each particle is evaluated using the fitness function defined as:
\begin{equation}
    f_{\rm fit}(\boldsymbol{\phi}^{(t)}_{s}) = \sum^{M}_{m=1}\sum^{K}_{k=1} R_{m,k}\left(\left\{\mathbf{F}_{{\rm A},m},\mathbf{F}_{{\rm D},m}\right\},\boldsymbol{\phi}^{(t)}_{s}\right).
\end{equation}
Notice that the objective function of problem (P6) is directly adopted as the fitness function, which effectively guides the search process toward improved sum-rate performance. Specifically, at each iteration, the fitness of all particles is computed, and both the individual and global best positions are
updated accordingly. This iterative procedure continues until the maximum iteration number $T$ is reached. The final solution to problem (P6), denoted by $\boldsymbol{\phi}^*$, is taken as the global best position, i.e., $\boldsymbol{\phi}^*=\boldsymbol{\phi}^{(T)}_{\rm g}$.

\begin{remark}(Algorithm convergence and computational complexity)
    \emph{First, we discuss the convergence of the proposed algorithm for solving problem (P1). For the inner-layer subproblem (P5), which updates $\mathbf{F}_{{\rm D},m}$, we note that its minimum value serves as an upper bound for the optimal value of problem (P4). By iteratively and optimally solving problem (P5), this upper bound is monotonically tightened. Consequently, the objective value of problem (P4) is non-increasing and converges to a stationary value \cite{zhang2024performance}. For the outer-layer subproblem, the PSO algorithm updates the variable $\boldsymbol{\phi}^{(t)}_{\rm g}$ across iterations such that $\boldsymbol{\phi}^{(t)}_{\rm g}\ge\boldsymbol{\phi}^{(t-1)}_{\rm g}$. Since the achievable sum-rate is upper-bounded, the convergence of the overall algorithm can be guaranteed.
Next, we analyze the computational complexity of the proposed algorithm, which is jointly determined by the inner optimization of $\mathbf{F}_{{\rm D},m}$ and the outer-layer update of $\boldsymbol{\phi}$. According to \cite{liu2025physical}, the complexity of the interior-point method for solving problem (P5) is on the order of $\mathcal{O}(K^6\log\frac{1}{\epsilon})$, where $\epsilon$ is the solution accuracy required by the CVX solver. Considering that the number of fitness evaluations in the PSO algorithm is $TS$, the total computational complexity of our algorithm is on the order of $\mathcal{O}(ITSK^6\log\frac{1}{\epsilon})$, where $I$ represents the number of SCA iterations.\footnote{The proposed double-layer algorithm, although computationally intensive, can be practically deployed under parallel computing architectures, since the fitness evaluations of different particles in the outer-layer PSO are mutually independent. Moreover, reduced-complexity designs can be achieved by adopting more efficient PSO encoding strategies for rotation angle optimization \cite{11197972} and by replacing the inner-layer SDR/SCA with lower-complexity beamforming schemes such as fractional programming (FP) or ZF. Real-time implementation with further complexity reduction is left for future work. }}
\end{remark}

\begin{remark}[Scalability and distributed implementation]
    \emph{The centralized optimization framework with global CSI and joint PSO-based rotation-angle design can, in principle, be extended to systems with a larger number of cells, but at the cost of increased computational complexity and signaling overhead due to the growing optimization dimension. In practical large-scale deployments, scalability can be improved by adopting a two-timescale architecture in which rotation angles, as slow-varying control variables, are optimized on a long timescale using large-scale or statistical channel information together with limited inter-cell message exchange, such as aggregated interference power or pricing coefficients, while digital beamforming is updated locally at each BS based on instantaneous CSI \cite{5608514}. Furthermore, the joint high-dimensional angle optimization can be replaced by distributed cell-wise or block-coordinate updates, which substantially reduce both computational complexity and signaling requirements. The development of more sophisticated and fully distributed coordination mechanisms tailored to large-scale mixed-field systems with rotatable arrays deserves further investigation and is left for future work.
    }
\end{remark}

\section{Numerical Results}\label{Sec:NR}
In this section, we present numerical results to demonstrate the benefits of the proposed RA-array scheme in enhancing the performance of multi-cell mixed-field communication systems, as well as the effectiveness of the proposed joint design.\footnote{Note that the simulations in this work do not account for the energy consumption associated with antenna array rotation. This omission stems from the absence of a well-established and accurate power consumption model for RA-enabled systems, as the rotation-related energy cost is highly dependent on specific hardware implementations \cite{zheng2025rotatableM}. A rigorous modeling of the mechanical energy consumption induced by array rotation, together with a comprehensive energy-efficiency analysis, is therefore beyond the scope of this paper and left for future investigation. Nevertheless, it is important to emphasize that the performance gains enabled by antenna rotation are inherently accompanied by additional mechanical energy expenditure, and the resulting trade-off should be carefully examined in the design of energy-efficient RA-enabled communication systems \cite{zheng2025rotatableJ}.}
\subsection{System Setup and Benchmark Schemes}
We consider the following system setup. Each BS in all cells is equipped with $N=129$ antennas and serves $K=3$ users, operating at a carrier frequency of $f=28$ GHz. In each cell, the $K$ near-field users are uniformly distributed within a region defined by a distance range of
$[0.1R_{\rm Ray},R_{\rm Ray}]$ and a spatial angle range of $[\frac{\pi}{3},\frac{2\pi}{3}]$. Unless otherwise specified, other system parameters are set as $P=30$ dBm, $\sigma^2_{m,k}=-80$ dBm \cite{11305158}, $L_{m,k}=3$, $\forall{m,k}$, $S=50$, $T=50$, $c_1=c_2=1.4$, $\omega_{\rm min}=0.4$, $\omega_{\rm max}=0.9$ \cite{11018493}, and $\mathcal{C}_{\phi_{m}}=[-\frac{\pi}{6},\frac{\pi}{6}]$, $\forall{m}$ \cite{zheng2025rotatableJ}.

 \begin{figure}[t]
	\centering
\includegraphics[width=0.45\textwidth]{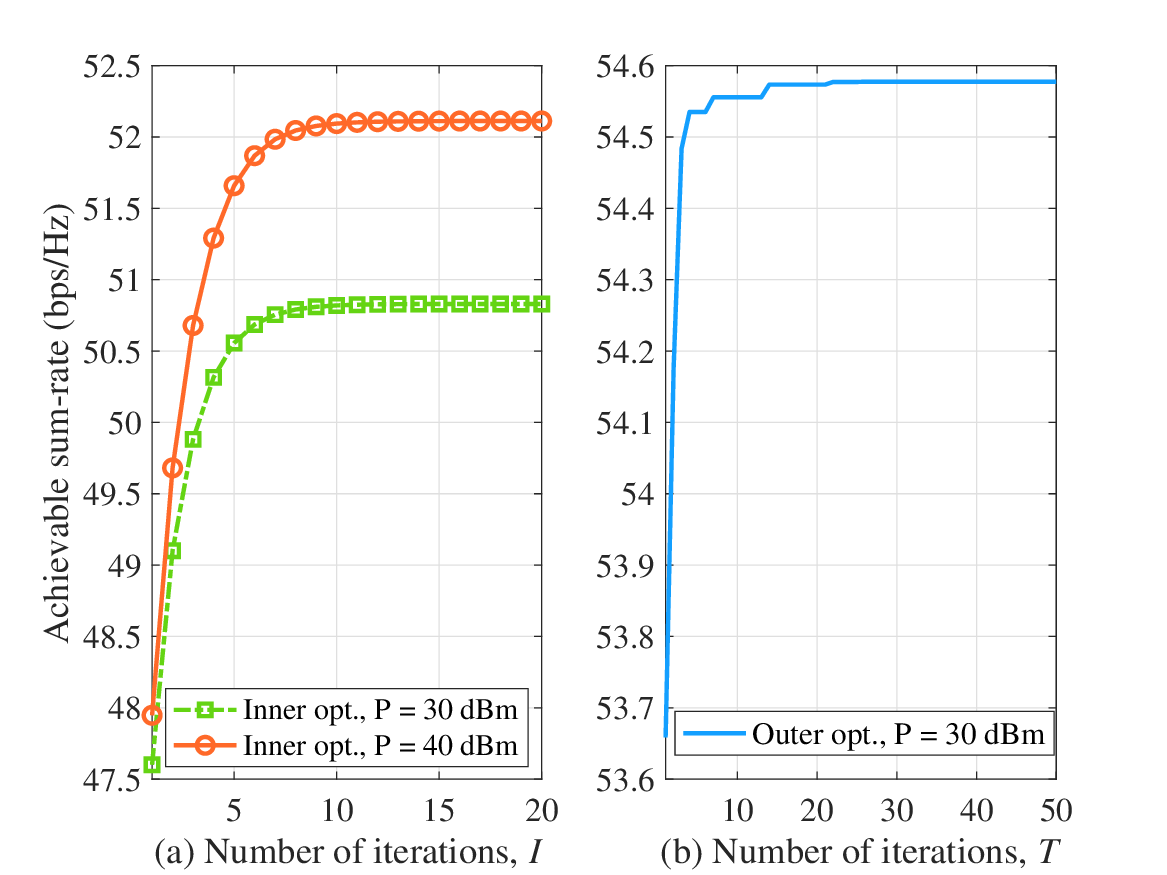}
	\caption{{Convergence of the proposed double-layer algorithm.}} \label{Fig:converagence}
\end{figure}
    \begin{figure}[t]
	\centering
\includegraphics[width=0.45\textwidth]{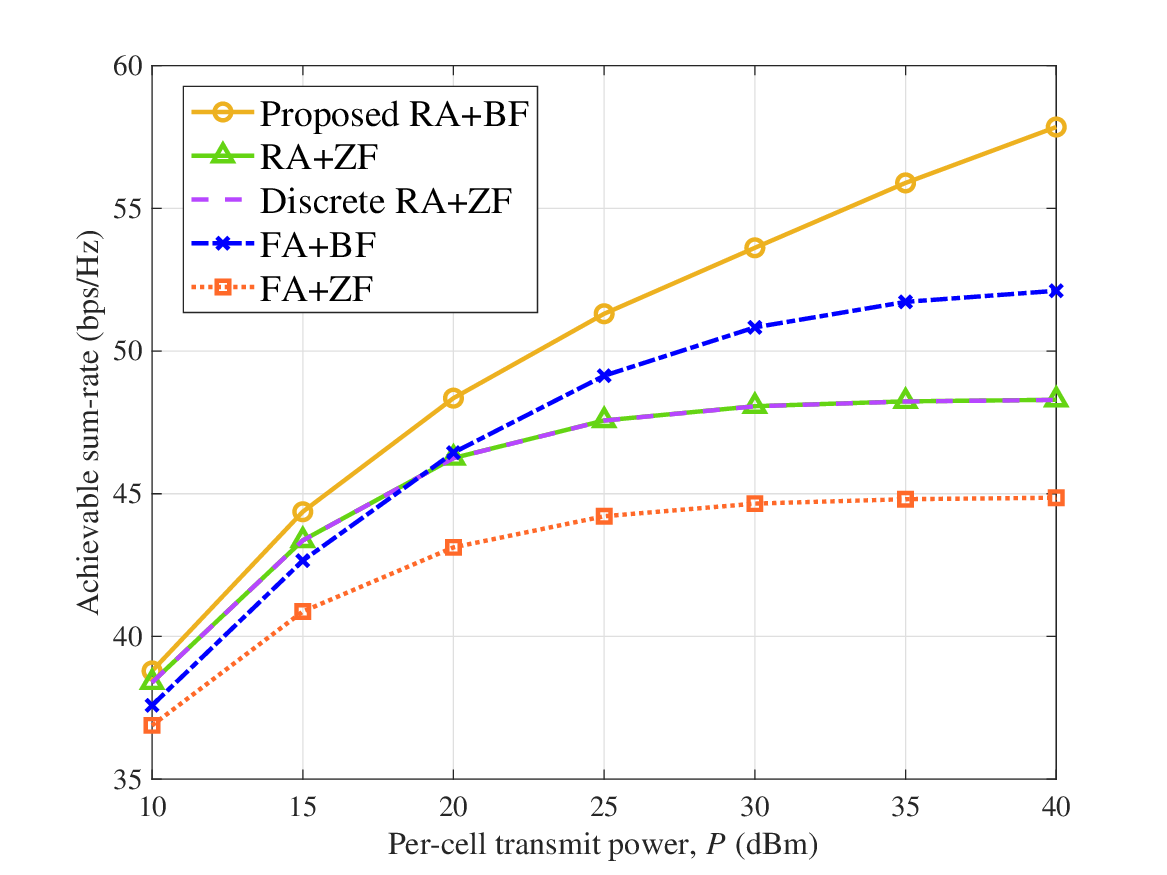}
	\caption{{Achievable sum-rate versus per-cell transmit power.}} \label{Fig:transpow}   
\end{figure}
 
We employ the following benchmark schemes for performance comparison:
\begin{itemize}
    \item \emph{Fixed Antenna + Beamformer Optimization (FA+BF):} In this scheme, all RA arrays are deactivated (i.e., the antenna angles are fixed), while the transmit beamforming matrix at each BS is optimized using the proposed method, as described in Section \ref{Sec:BF_opt}. 
    \item \emph{Fixed Antenna + ZF (FA+ZF):} In this scheme, all RA arrays are deactivated (i.e., the antenna angles are fixed), while the transmit beamforming matrix at each BS is optimized using the ZF. 
    \item \emph{Rotation Optimization + ZF (RA+ZF):} The rotation angle vector is optimized using the proposed method in Section \ref{Sec:Rot_Opt}, while the transmit beamformer is designed based on ZF.
    \item \emph{Discrete Rotation Angle Optimization + ZF (Discrete RA+ZF):} In this scheme, the rotation angle of each RA array is uniformly discretized into $100$ points. The transmit beamformer is then designed using ZF. All possible combinations of rotation angles are exhaustively enumerated, and the combination yielding the highest achievable rate is selected as its performance.
\end{itemize}
  
    \begin{figure}[t]
	\centering
\includegraphics[width=0.45\textwidth]{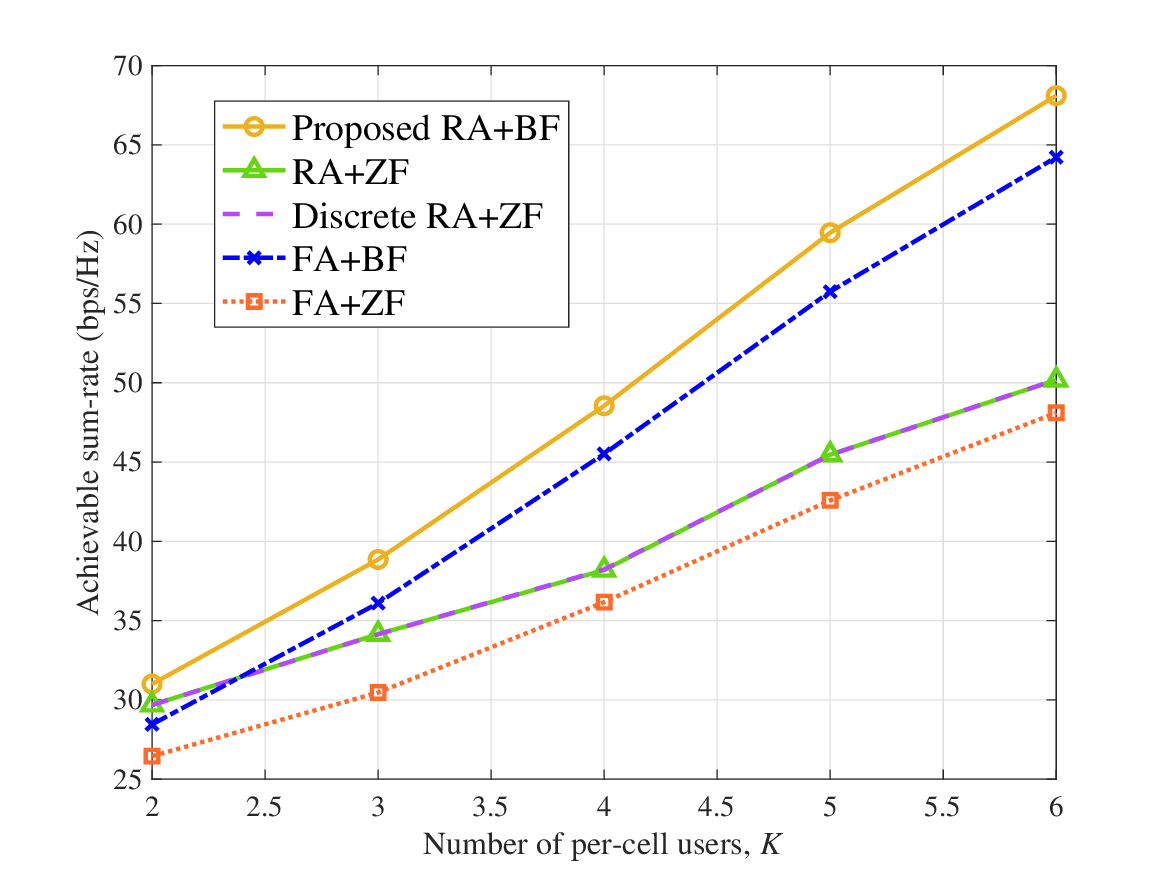}
	\caption{{Achievable sum-rate versus number of per-cell users.}} \label{Fig:numUsers}   
\end{figure}
\subsection{Convergence of the Proposed Algorithm}
In Fig. \ref{Fig:converagence}, we plot the convergence behavior of the proposed beamformer design and the PSO algorithm for solving problem (P1). As shown in Fig. \ref{Fig:converagence}(a), the achievable sum-rate of the proposed beamforming algorithm converges to a stationary point after approximately $15$ iterations. Meanwhile, Fig. \ref{Fig:converagence}(b) demonstrates that the PSO algorithm reaches a stationary point within about $30$ PSO iterations.
    \begin{figure}[t]
	\centering
\includegraphics[width=0.45\textwidth]{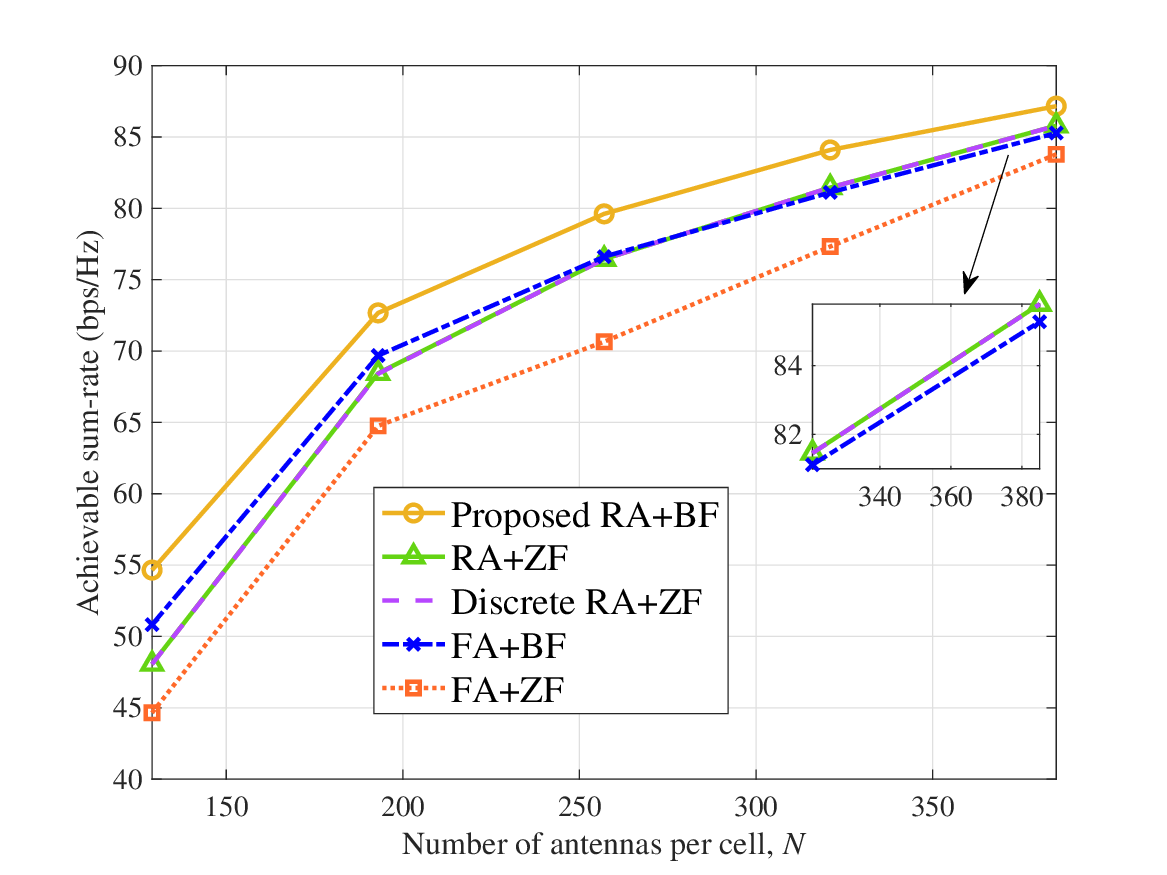}
	\caption{{Achievable sum-rate versus number of per-BS antennas.}} \label{Fig:numAntenna}    
\end{figure}

\subsection{Effect of Per-cell Transmit Power}
In Fig. \ref{Fig:transpow}, we plot the achievable sum-rate of users across all cells versus per-cell transmit power. The devised algorithm consistently outperforms all benchmark schemes, with the performance advantage becoming more pronounced at a higher transmit power ($P\ge30$ dBm). All methods show a monotonic increase in sum-rate with the transmit power. Nevertheless, except for our approach, the benchmark schemes exhibit saturation, leading to an increasing performance gap. The reasons are provided as follows. For the RA+ZF scheme, the array rotation must be appropriately determined to balance two competing factors: reducing the array loss caused by ZF-based transmit beamformer and suppressing inter-cell mixed-field interference. This trade-off limits its overall performance. For the FA+ZF scheme, although ZF-based beamforming effectively mitigates the intra-cell near-field interference, it comes at the expense of array gain, and the inter-cell mixed-field interference, however, cannot be suppressed using ZF, thereby degrading the system performance. For FA+BF, the lack of array rotation reduces the transmit beamforming capability for mitigating both the intra-cell near-field interference and the inter-cell mixed-field interference.  Consequently, it becomes less effective than the proposed scheme, thus rendering the effectiveness of the RA-enabled multi-cell system with joint transmit beamforming and array rotation design.
Moreover, the RA+ZF scheme outperforms FA+BF when $P\le 20$ dBm, but becomes inferior when the transmit power exceeds this threshold. This is because, at lower power levels, where intra-cell mixed-field interference is moderate, the combination of array rotation and ZF beamforming can effectively handle both near-field and mixed-field interference. However, when mixed-field interference becomes more severe at higher power levels, mitigating it requires more sophisticated beamforming strategies.
Finally, RA+ZF achieves nearly the same performance as the discrete RA+ZF method, confirming the effectiveness of the developed rotation-angle optimization algorithm.
\subsection{Effect of Number of Per-cell Users}
In Fig. \ref{Fig:numUsers}, we plot the achievable sum-rate of users across all cells versus number of users per cell. The near-field users in each cell are sequentially sampled from a region defined by a distance range of $[0.1R_{\rm Ray},R_{\rm Ray}]$ and a spatial angle range of $[\frac{\pi}{3},\frac{2\pi}{3}]$. We first observe that the proposed approach significantly outperforms the benchmark schemes, and that all of them exhibit a monotonic increase in sum-rate as the number of users per cell increases. This trend arises because, although complex intra-cell near-field and inter-cell mixed-field interference exist, the achievable sum-rate is primarily determined by the rates of near-field users in each cell, leading to higher overall rates as the user number grows. Second, the growth rate attained by the proposed RA+BF and FA+BF methods is substantially higher than that of the ZF-based schemes. This implies that, despite the increasingly severe intra-cell near-field and inter-cell mixed-field interference with larger $K$, the transmit beamformer design presented in Section \ref{Sec:BF_opt} demonstrates a stronger capability to effectively mitigate such interference. In addition, the RA+ZF scheme outperforms the FA+BF approach when $K=2$, highlighting the effectiveness of combining array rotation with ZF-based beamformer under moderate interference conditions.

\subsection{Effect of Number of Antennas Per BS}
In Fig. \ref{Fig:numAntenna}, we present the achievable sum-rate of different schemes versus the number of antennas per BS. As expected, the sum-rates of all schemes increase with the antenna number, attributed to the improved beamforming gain. The proposed joint design consistently outperforms all benchmark schemes across different antenna settings. An interesting observation is that the FA+BF method initially achieves higher performance than the RA+ZF scheme, but the gap gradually narrows as the number of antennas increases. Once the number of antennas exceeds $N \ge 257$, the former scheme becomes inferior to the latter. This is because a larger antenna array provides finer spatial resolution, allowing ZF-based beamformer to more effectively suppress intra-cell interference while causing only moderate inter-cell interference. As a result, RA-assisted ZF beamforming ultimately delivers superior rate performance compared to FA+BF.


\section{Conclusions}\label{Sec:con}
In this paper, we proposed deploying RA arrays at each BS to enhance multi-cell communication performance by exploiting the additional spatial DoFs offered by RA array to mitigate intra-cell near-field interference and inter-cell mixed-field interference. For the special case with a single user per cell, we derived the closed-form expression for the inter-cell mixed-field interference based on the Fresnel integrals and identified the key factors influencing such interference, which is in sharp contrast to conventional single-cell scenarios.
We further demonstrated that array rotation can effectively suppress complex inter-cell mixed-field interference, leading to significant performance gains. For the general scenario with multiple users per cell, we designed a double-layer optimization framework, with the inner layer optimizing the hybrid beamforming and the outer layer determining the rotation angles across all BSs. Numerical results verified the superiority of the devised RA-enabled multi-cell communication system over the fixed-antenna approaches. 

\begin{appendices}
    \section{Proof of Proposition \ref{The:NF_inter}}\label{App0}
    Let $A_1=\sqrt{\frac{d\sin^2{(\phi_i-\theta_{i,m})}}{2r_{i,m}}}$ and $A_2=\frac{\cos{(\psi_{i,k}-\phi_i)}- \cos{(\phi_i-\theta_{i,m})}}{2A_1}$. Using the Riemann integral, we rewrite \eqref{Eq:rho_unified} as $\rho(\psi_{i,k},\theta_{i,m},r_{i,m},\phi_i)\approx\frac{1}{N}\left|\int^{\tilde{N}}_{-\tilde{N}}e^{j\pi(A_1n+A_2)^2}dn\right|$ \cite{zhang2023mixed,zhang2023swipt}. With the substitution $t = \sqrt{2}A_1n+\sqrt{2}A_2$, we obtain $\rho(\psi_{i,k},\theta_{i,m},r_{i,m},\phi_i)=\frac{1}{\sqrt{2}A_1N}\left|\int^{\sqrt{2}A_1\tilde{N}+\sqrt{2}A_2}_{-\sqrt{2}A_1\tilde{N}+\sqrt{2}A_2}e^{j\pi\frac{1}{2}t^2}dt\right|$.
    We define   \begin{align}
        \gamma^{(1)}_{i,k,m} &=(\cos{(\phi_i\!-\!\theta_{i,m})}\!-\!\cos{(\psi_{i,k}\!-\!\phi_i)})\sqrt{\frac{r_{i,m}}{d\sin^2{(\phi_i\!-\!\theta_{i,m})}}},\\
        \gamma^{(2)}_{i,m} &=\frac{N}{2} \sqrt{\frac{d\sin^2{(\phi_i-\theta_{i,m})}}{r_{i,m}}}.
    \end{align}
    Then, we have 
\begin{align}\label{Eq:rho_unified_approx_pro}
&\rho\left(\psi_{i,k},\theta_{i,m},r_{i,m},\phi_i\right) \approx G\left(\gamma^{(1)}_{i,k,m},\gamma^{(2)}_{i,m}\right) \nn\\ &\quad\quad\quad\quad=\left|\frac{\widehat{C}\left(\gamma^{(1)}_{i,k,m},\gamma^{(2)}_{i,m}\right)+j\widehat{S}\left(\gamma^{(1)}_{i,k,m},\gamma^{(2)}_{i,m}\right)}{2\gamma^{(2)}_{i,m}}\right|,
    \end{align}
    where $\widehat{C}(\gamma_1,\gamma_2)=C(\gamma_1+\gamma_2)-C(\gamma_1-\gamma_2)$ and $\widehat{S}(\gamma_1,\gamma_2)=S(\gamma_1+\gamma_2)-S(\gamma_1-\gamma_2)$. Note that, the general expression in \eqref{Eq:rho_unified_approx_pro} reduces to the specific interference terms $\rho(\psi_{2,1},\theta_{2,1},r_{2,1},\phi_2)$ and $\rho(\psi_{1,2},\theta_{1,1},r_{1,1},\phi_1)$ by setting the indices $(i,k,m)$ to $(2,1,1)$ and $(1,2,1)$, respectively. This completes the proof.
      \section{Proof of Corollary \ref{Pro:equal}}\label{App2}
      We detail the proof for the following three cases.

\textbf{Case 1:} When $\psi_{2,1}=\theta_{2,1}=\frac{\pi}{2}$, the normalized inter-cell mixed-field interference depends only on $\gamma_2$, i.e., $\rho(\psi_{2,1},\theta_{2,1},r_{2,1},\phi_2)\approx G(0,\gamma_2)$,
where
$ \gamma_2 =\frac{N}{2} \sqrt{\frac{d\cos^2{\phi_2}}{r_{2,1}}}.
$
Since $G(0,\gamma_2)$ generally decreases with $\gamma_2$, the optimal rotation angle $\phi_2$ that minimizes $G(0,\gamma_2)$ is the one that maximizes $\gamma_2$, which occurs at $\phi_2=0$. Thus, the suboptimal solution to problem (P2) is $\phi_2^*=0$.

\textbf{Case 2:} When $\psi_{2,1}=\theta_{2,1}\neq\frac{\pi}{2}$,
the normalized inter-cell mixed-field interference also depends only on $\gamma_2$, i.e., $\rho(\psi_{2,1},\theta_{2,1},r_{2,1},\phi_2)\approx G(0,\gamma_2)$,
where
$
    \gamma_2 =\frac{N}{2} \sqrt{\frac{d\sin^2{(\phi_2-\theta_{2,1})}}{r_{2,1}}}.
$
To reduce the interference, $\phi_2$ should be chosen to maximize $\gamma_2$, whose maximum is attained at $|\phi_2-\theta_{2,1}|=\frac{\pi}{2}$. Considering the admissible rotation region of $\phi_2$, if $\theta_{2,1}>\frac{\pi}{2}$, the suboptimal rotation angle is $\phi_2^*= \max(\frac{\pi}{2}-\psi_{2,1},\phi_{2,{\rm min}})$, whereas if $\theta_{2,1}<\frac{\pi}{2}$, $\phi_2^*= \min(\frac{\pi}{2}-\psi_{2,1},\phi_{2,{\rm max}})$. 

\textbf{Case 3:} When $\psi_{2,1}\neq\theta_{2,1}$,
we first consider the case when $\psi_{2,1}<\theta_{2,1}$, with $\psi_{2,1},\theta_{2,1}\in (0,\frac{\pi}{2})$.
Define the following functions:
 \begin{align}
        &\Delta_1(\phi_2)= \nn\\   
        &\frac{|\cos{(\phi_2-\theta_{2,1})}-\cos{(\psi_{2,1}-\phi_2)}|}{|\sin{(\phi_2-\theta_{2,1})}|} - \frac{|\cos{\theta_{2,1}}-\cos{\psi_{2,1}}|}{|\sin{\theta_{2,1}}|},\nn\\
        &\overset{(a_1)}{=}\frac{\cos{(\psi_{2,1}-\phi_2)}-\cos{(\theta_{2,1}-\phi_2)}}{\sin{(\theta_{2,1}-\phi_2)}} - \frac{\cos{\psi_{2,1}}-\cos{\theta_{2,1}}}{\sin{\theta_{2,1}}},
    \end{align}
    and
  \begin{align}
        \Delta_2(\phi_2) &=  |\sin{(\phi_2-\theta_{2,1})}|-|\sin{\theta_{2,1}}|\nn\\
        &\overset{(a_2)}{=}\sin{(\theta_{2,1}-\phi_2)}-\sin{\theta_{2,1}},
    \end{align}
where $(a_1)$ uses the fact that the cosine function is monotonically decreasing over $(0,\frac{\pi}{2})$, and $(a_2)$ holds since $\phi_2<\theta_{2,1}$. Moreover, it is straightforward to verify that $\Delta_1(0)=\Delta_2(0)=0.$ 
To show the existence of a feasible $\phi_2$ such that $\Delta_1(\phi_2)>0$ and $\Delta_2(\phi_2)>0$, we analyze their Taylor expansions around $\phi_2=0$. The first-order derivatives with respect to $\phi_2$, respectively, are:
\begin{align}
    \Delta^{\prime}_1(\phi_2)&= \frac{\cos(\theta_{2,1}-\psi_{2,1})-1}{\sin^2{(\theta_{2,1}-\phi_2)}} < 0,\nn\\
    \Delta^{\prime}_2(\phi_2)&= -\cos{(\theta_{2,1}-\phi_2)}< 0.
\end{align}

Specifically, we expand $\Delta_2(\phi_2)$, using Taylor’s theorem:
\begin{align}
    \Delta_2(\phi_2) =& \Delta_2(0)+\Delta^{\prime}_2(\phi_2)\phi_2 + \mathcal{O}(\phi^2_2)\nn\\
    =
    &-\cos{(\theta_{2,1})}\phi_2 + \mathcal{O}(\phi^2_2).
\end{align}
Since $-\cos{(\theta_{2,1})}<0$, choosing $\phi_2 = -\epsilon$, with sufficiently small $\epsilon>0$, $\Delta_2(\phi_2)>0$ is ensured.
Similarly, the Taylor expansion of $ \Delta_1(\phi_2)$ is given by
\begin{align}
    \Delta_1(\phi_2) &=  \Delta_1(0)+\Delta^{\prime}_1(\phi_2)\phi_2 + \mathcal{O}(\phi^2_2)\nn\\
    &=
    \frac{\cos(\theta_{2,1}-\psi_{2,1})-1}{\sin^2{(\theta_{2,1}-\phi_2)}}\phi_2 + \mathcal{O}(\phi^2_2).
\end{align}
Since $\Delta^{\prime}_1(\phi_2)<0$ and $\phi_2<0$, it follows that  $\Delta_1(\phi_2)>0$. Thus, by setting $\phi_2 = -\epsilon$, both $\Delta_1(\phi_2)>0$ and $\Delta_2(\phi_2)>0$ are satisfied. 
Notice that for the case $\psi_{2,1}>\theta_{2,1}$,  the argument is similar, with sign changes accordingly. 
Furthermore, when $\psi_{2,1},\theta_{2,1}\in(\frac{\pi}{2},\pi)$, the same proof applies due to symmetry in the trigonometric functions. This completes the proof.
\section{}\label{App3}
When $\psi_{2,1}=\theta_{2,1}$, the inter-cell mixed-field interference depends only on $\gamma_2$, i.e., $\rho(\psi_{2,1},\theta_{2,1},r_{2,1},\phi_2)\approx G(0,\gamma_2)$,
where
$\gamma_2 =\frac{N}{2} \sqrt{\frac{d\sin^2{(\phi_2-\theta_{2,1})}}{r_{2,1}}}.$
Specifically, as $r_{2,1}$ increases, $\gamma_2$ decreases, which in turn causes $\rho(\psi_{2,1},\theta_{2,1},r_{2,1},\phi_2)$ to increase, thereby reducing the rate.
In addition, it is straightforward to verify that the rotation angle $\phi_2$ can be chosen such that $\sin^2{(\phi_2-\theta_{2,1})}$ attains its maximum value of $1$. However, since the range of ${\rm U}_{2,1}$ is much greater than the maximum attainable value of $\sin^2{(\phi_2-\theta_{2,1})}$, the effect of adjusting $\phi_2$ to suppress the interference becomes nearly negligible with increasing $r_{2,1}$, especially at sufficiently large distances.
\end{appendices}
\bibliographystyle{IEEEtran}
\bibliography{IEEEabrv,Ref}

\end{document}